\documentclass[preprint,prc,a4paper,tightenlines,nofootinbib]{revtex4}
\usepackage{amssymb}
\usepackage{amsmath}
\usepackage{graphicx}
\usepackage{bm} % for bold greek letters e.g. \bm{\tau}

\usepackage{xcolor}

\usepackage[active]{srcltx}

\newcommand{\dis}{\displaystyle}
\newcommand{\taup}{\tau_{\!+}}

\newcommand{\taux}{\tau_{\!\times}}

\newcommand{\maM}{\mathcal{M}}

\newcommand{\boldpi}{\bm{\pi}}
\newcommand{\boldtau}{\bm{\tau}}
\newcommand{\boldphi}{\bm{\phi}}

\newcommand{\mpi}{\ensuremath{m_\pi}}   % pion mass
\newcommand{\mN}{\ensuremath{m_{N}}}   % nucleon mass
\newcommand{\gA}{\ensuremath{g_{\!A}}} % NNpi coupling constant
\newcommand{\fpi}{\ensuremath{f_{\!\pi}}} % pion decay constant 92 MeV
\newcommand{\nolik}{\ensuremath{0}}    % zero or epsilon for propagators i.e. +i0
\newcommand{\NNNNpi}{\ensuremath{NN \to NN\pi}}% N N -> N N pi
\newcommand{\NLO}{NLO}
\newcommand{\NNLO}{N$^2$LO} % please, use \NNLO{} to avoid problems with spaces
\newcommand{\NNNNLO}{N$^4$LO}

\newcommand{\vdotq}{\ensuremath{v \cdot q }}

\newcommand{\ipipi}{{I_{\pi \pi}}}
\newcommand{\ipipifsi}{{I_{\pi \pi}^{\rm finite}}}%^{finite, scaleindep}}}

\newcommand{\jpid}{{J_{\pi  \Delta }}}
\newcommand{\jpidfsi}{{J_{\pi  \Delta }^{\rm finite}}}%^{finite, scaleindep}}}
\newcommand{\jpipin}{{J_{\pi \pi N}}}
\newcommand{\jpipid}{{J_{\pi \pi \Delta}}}
\newcommand{\jpipind}{{J_{\pi \pi N \Delta}}}

\newcommand{\intI}{\ipipi}
\newcommand{\intJ}{\frac{\jpid}{\delta}}
\newcommand{\intT}{\delta \jpipid}
\newcommand{\intC}[1]{\frac{#1}{(4\pi)^2}}
\newcommand{\intL}{{k_1^2 \jpipind}}

\newcommand{\ga}{g_{A}}
\newcommand{\gpind}{g_{\pi N \Delta}}
\newcommand{\gone}{g_{1}}

\newcommand{\Sd}{\mathbb{S}}
\newcommand{\Sdd}{\mathbb{S}^{\dagger}}

\begin{document}

\title{Pion production in nucleon-nucleon collisions in chiral
  effective field theory with  Delta(1232)-degrees  of  freedom.     }

\author{A.~A.~Filin}
\affiliation{Institut f\"ur Theoretische Physik II, Ruhr-Universit\"at 
Bochum, D-44780 Bochum, Germany}
\affiliation{Institute for Theoretical and Experimental Physics, 117218, 
B.~Cheremushkinskaya 25, Moscow, Russia}

\author{V.~Baru}
\affiliation{Institut f\"ur Theoretische Physik II, Ruhr-Universit\"at 
Bochum, D-44780 Bochum, Germany}
\affiliation{Institute for Theoretical and Experimental Physics,
 117218, B.~Cheremushkinskaya 25, Moscow, Russia}

\author{E.~Epelbaum }
\affiliation{Institut f\"ur Theoretische Physik II, Ruhr-Universit\"at 
Bochum, D-44780 Bochum, Germany}

\author{C.~Hanhart}
\affiliation{Institut f\"{u}r Kernphysik,  (Theorie) and J\"ulich Center for Hadron Physics,
 Forschungszentrum J\"ulich,  D-52425 J\"{u}lich, Germany and\\
Institute for Advanced Simulation, Forschungszentrum J\"ulich,  D-52425 J\"{u}lich, Germany}

\author{H.~Krebs}
\affiliation{Institut f\"ur Theoretische Physik II, Ruhr-Universit\"at Bochum, 
D-44780 Bochum, Germany}

\author{F.~Myhrer}
\affiliation{Department of Physics and Astronomy, University of South Carolina,
Columbia, SC 29208, USA}

\begin{abstract} 
A   calculation of  the pion-production operator up to next-to-next-to-leading order  
for  s-wave pions is performed  
within chiral effective field theory.   
In the previous study   [Phys.\ Rev.\ C 85, 054001 (2012)] we  discussed  
the  contribution of the pion-nucleon loops  at  the same order.  
Here  we  extend  that  study to  include    
explicit Delta degrees  of freedom and  the $1/m_N^2$ corrections  to  the  
pion-production amplitude.  
Using  the power counting scheme  where  the  Delta-nucleon mass  
difference  is  of   the  order of  the  characteristic 
momentum scale  in the  production process,  we  calculate  all  
tree-level  and  loop  diagrams   involving Delta up  to 
next-to-next-to-leading order.    
The  long-range part 
of the Delta loop  contributions  is  found   to  be  of  similar  size  to 
that   from  the  pion-nucleon  loops which supports 
  the  counting  scheme.  The  net  effect of pion-nucleon  and  
Delta  loops  is expected to   play  a  crucial role  in  
  understanding  of   the  neutral  pion  production  data.

\end{abstract}

\maketitle

%%%%%%%%%%%%%%%%%%%%%%%%%%%%%%%%%%%%%%%%%%%%%%%%%%%%%%%%%%%%%%%%%%%%%%%%%%%%%%%%%%%
\section{INTRODUCTION}
%%%%%%%%%%%%%%%%%%%%%%%%%%%%%%%%%%%%%%%%%%%%%%%%%%%%%%%%%%%%%%%%%%%%%%%%%%%%%%%%%%% 

The reaction $NN\to NN\pi$, being the first inelastic channel of $NN$
interaction,   provides a good  possibility  to study  $NN$
dynamics at intermediate energies.   
The interest in pion production
in $NN$ collisions  was revived  almost 20  years ago  when it was
proposed  \cite{cohen,park},    
that the process may be studied in a model-independent way
within   chiral perturbation theory  (ChPT)  --- the low-energy
effective field theory of QCD ---  lately reviewed in Refs.~\cite{hanhart04,BHM}. 
In this effective field theory the approximate chiral symmetry of QCD 
is exploited in a systematic way. 
The study of the $NN\to NN\pi$ reaction at threshold builds on the 
detailed understanding of the necessary two-body ingredients, the 
$\pi N$ and $NN$ interactions and reactions, 
which have   been  successfully evaluated  within ChPT
at low energies, see, e.g.,  Refs.~\cite{Bernardrev} 
and \cite{NN3}  for recent reviews.

Surprisingly, the naive  application  of the  standard ChPT  power counting,   
where one assumes the  typical  momenta  $|\vec p\, |$ 
to be of the order of the pion mass $\mpi$,  
to  the reaction $NN\to NN\pi$ by  Refs.~\cite{park,sato,unserd,DKMS,Ando,novel},   
did not only  fail to  describe  the data, especially 
in the $pp\to pp\pi^0$ channel, 
but also revealed a problem  with the  convergence of the chiral  expansion. 
However, in  Ref.~\cite{cohen,rocha}  
it  was  advocated that the proper treatment of the   reaction  $NN\to NN\pi$  requires taking into account 
the new intermediate momentum scale,   $p=|\vec p\, | \sim \sqrt{\mpi m_N}$  
($m_N$ is the nucleon mass), which corresponds to 
the  relative  momentum  of the initial  nucleons required to produce a pion  at  threshold.  
This  new scale led  to  a new  hierarchy  of   diagrams   driven by the   expansion parameter 
\begin{equation}
\chi_{\rm MCS}\simeq\frac{\mpi}{p}\simeq\frac{p}{\Lambda_\chi}\simeq 
\sqrt{\frac{\mpi}{m_N}},
\label{expansionpapar}
\end{equation}
with $\Lambda_\chi \simeq m_N$ being the typical hadronic scale. 
In  what  follows,  we will refer  to  this  power counting  as  the  
momentum  counting  scheme  (MCS).  
The  MCS  has  been  applied    in  a  variety  of  pion threshold production reactions  for the 
outgoing pion in an  s-wave  \cite{HanKai,towards,ksmk09,NNLOswave,subloops,BM} and 
p-wave \cite{ch3body,newpwave,nakamura}, and for 
isospin violating  pion  production reactions \cite{KNM,weCSB,Bolton}.

It  has  been  known for  years  that  the strength  of  the s-wave  
pion  production amplitude  
in the  charged  channels $pp\to d\pi^+$ and 
$pp\to pn\pi^+$   is dominated  by the  leading  order (LO) 
Weinberg--Tomozawa (WT) operator \cite{koltun}.   
On  a  more   quantitative 
level, the  cross  sections  for these two reactions were  underestimated  
by  a  factor   of  2 \cite{hanhart04}. 
Meanwhile, the  application  of  the  MCS  to  s-wave  production  in  the  
$pp\to d\pi^+$ channel  at  next-to-leading order  (NLO) \cite{towards},
revealed quite  good  agreement   with  experimental  data.     
To obtain this  agreement it  was  important  to realize  that  the  nucleon recoil corrections 
$\propto 1/m_N$ contribute  in MCS    at  lower  order  than  what  is  
indicated naively by the order of the  Lagrangian  since in MCS $p^2/m_N \sim m_\pi$.   
In effect, the  leading   WT   operator  was enhanced  by  a  factor  4/3  
due to  the  recoil correction   to  the WT  pion rescattering vertex.
This enhancement resulted in an increase  
of  the cross section  by  about the missing  factor 2.  
In contrast, for  s-wave pion production in  the  neutral  channel  $pp\to pp\pi^0$, the 
situation  is completely  different.
In this channel  the  large isovector WT  rescattering vertex  does not  contribute while
the  direct  emission of  the pion  from  one-nucleon  leg at LO  is  
dynamically strongly  suppressed~\cite{cohen}.    
Furthermore,  the  resulting  contribution   of loops  at  NLO was  shown  to  vanish  
both for the  $pp\to pp\pi^0$  \cite{HanKai}  and  
$pp\to d\pi^+$   \cite{towards}   reaction channels. 
In fact, we  believe  that the experimentally measured  $pp\to pp \pi^0$ reaction  
is unique in that it directly probes 
the higher order MCS contributions which in the charged channels 
are masked by the dominant lower 
order Weinberg--Tomozawa  term.  
Following  this logic, 
in  Ref.~\cite{NNLOswave}  we  extended  the  analysis  of 
pion-nucleon loops   
to next-to-next-to-leading order (\NNLO{}).   
The pertinent results of  Ref. \cite{NNLOswave}  
can be  summarized as  follows: 
%%%%%%%%%%%%%%%
\begin{itemize}
\item   significant  cancellations  of loops  found at  NLO
are  also  operative  at \NNLO{}.  In  particular,  
all  loop  topologies  involving  $1/m_N$  corrections  to  
the leading  vertices cancel  completely,  as  do the  
 loops  involving low-energy  constants (LECs)  $c_i$. 
\item  the  cancellation of pion-nucleon loops  at  \NNLO{} is  not  complete  
yielding  a  non-vanishing  \NNLO{} contribution.  
\item using dimensional regularization, all UV divergencies in the loops were absorbed
into redefinition of  low-energy-constants (LECs)  in the Lagrangian at  \NNLO{}. 
These LECs parametrize  short-range physics not resolved  explicitly 
at the energies relevant for pion production at threshold.
\item  
the  scheme-independent  long-range  part  of  pion-nucleon loops  
was  found  to  be   of  a natural  size as  expected  from the MCS, 
both  for   charged  and neutral  pion  production. 
This puts in  question the earlier phenomenological 
studies~\cite{Lee,HGM,Hpipl,jounicomment,eulogio,unsers,unserdelta} of these reactions 
where none of these \NNLO{} loop contributions were 
considered. 
\end{itemize}
%%%%%%%%%%%%%

Due to the fact that the Delta-nucleon ($\Delta$-N) mass splitting, $\delta=m_\Delta - m_N$,
is numerically of the order of $p$ (i.e. $\delta \sim p$), the Delta-isobar should be 
explicitly included as a dynamical degree of freedom\footnote{In Ref.\cite{compton} the  same power  counting was applied to
study  Compton scattering off the  nucleon in the delta region}~\cite{cohen,rocha,HanKai}.   
At  tree-level, the  effect  of the Delta  manifests  itself   already 
at  NLO as discussed within chiral EFT in
 Refs.~\cite{cohen,rocha,HanKai,NNpiMenu}. 
In  addition,   starting from NLO  the  Delta-resonance  also gives  rise to  loop  contributions  
which  at NLO  were  shown in Ref.~\cite{HanKai} to  cancel  exactly 
both  for the  neutral  and  charged  channels.  
Meanwhile,  the  role  of the  Delta  loops  at  \NNLO{}  has  not  
been discussed  in the literature 
and is a topic of the present investigation.  

In  this  work  we  complete  the  calculation  of  the loop diagrams for 
the s-wave pion-production operator  at  \NNLO{}.   
In particular,  the previous calculation~\cite{NNLOswave} 
is  extended and   improved 
 in  the  following  aspects: 
%%%%%%%%%%%%%%%
\begin{itemize}
\item   we  treat  the  Delta-resonance  as an  explicit  
virtual degree  of  freedom 
in all loops and confirm the   cancellation   of  all loop  
contributions  containing Delta  at  NLO.   
\item    we  extend the  calculation  of  the  loops with 
the explicit Delta to  \NNLO{}. 
It is found that some of the  loops  renormalize the  bare  
$\pi NN$  coupling constant 
while a group of other  \NNLO{} loop diagrams 
vanishes  in a close analogy  with pion-nucleon loops. 
\item the \NNLO{} remnant  of the loops yields  a long-range  
contribution to the pion-production operator amplitudes similar in size  
to  those  from  pion-nucleon 
 loops at the  same  order.   
\item   we  include  the $1/m_N^2$  corrections  to  the  tree-level  
operators  at  \NNLO{}   
and   present  the    operator  
  for  s-wave  pion production  with  explicit  Delta  degrees  of  
freedom up-to-and-including  \NNLO{}. 
\end{itemize}
%%%%%%%%%%%%% 

The paper is  organized  as  follows. 
In the next section we review the arguments and results of Ref.~\cite{NNLOswave}. 
In particular, in this section we present some features special to the MCS.
In Sec.~II we discuss the cancellation 
mechanism of NLO loop diagrams  before we 
explicitly give the expressions for the tree-level and 
loop-diagram production operators   to \NNLO{} derived in Ref.~\cite{NNLOswave}. 
In Sec.~III we present the Lagrangian which includes the $\Delta$, 
give the tree-level amplitudes for the $\Delta$ contribution to 
s-wave pion production and then 
evaluate all the loop diagrams involving Delta excitation to \NNLO{}.
Here we heavily rely on the arguments presented in Sec.~II. 
The last subsection in Sec.~III contains the details on  regularization 
of the UV divergences present in the loop diagrams including 
a discussion of the necessary decoupling requirements  
in the heavy $\Delta$ mass limit.
In Sec.~IV we compare the pion-nucleon and $\Delta$-loop contributions 
before we make our conclusions in the last section.

%%%%%%%%%%%%%%%%%%%%%%%%%%%%%%%%%%%%%%%%%%%%%%%%%%%%%%%%%%%%%%%%%%%%%%%%%%%%%%%%
\section{S-WAVE PION PRODUCTION TO \NNLO{}: THE CONTRIBUTIONS OF PION AND NUCLEON DEGREES OF FREEDOM} 
\label{sec:nucl}
%%%%%%%%%%%%%%%%%%%%%%%%%%%%%%%%%%%%%%%%%%%%%%%%%%%%%%%%%%%%%%%%%%%%%%%%%%%%%%%
 
 For completeness and to prepare the stage for the main part of the
 paper, where the inclusion of the $\Delta$ as dynamical degree of freedom
 is discussed, in this section we   summarize the results of Ref.~\cite{NNLOswave}.

%%%%%%%%%%%%%%%%%%%%%%%%%%%%%%%%%%%%%%%%%%%%%%%%%%%%%%%%%%%%%%%%%%%%%%%%%%%%%%
\subsection{Diagrams and Power Counting}
\label{sec:PC}
%%%%%%%%%%%%%%%%%%%%%%%%%%%%%%%%%%%%%%%%%%%%%%%%%%%%%%%%%%%%%%%%%%%%%%%%%%%%%%%%%%%%%%%%%%%

\begin{figure}[h]
\includegraphics[width=13cm]{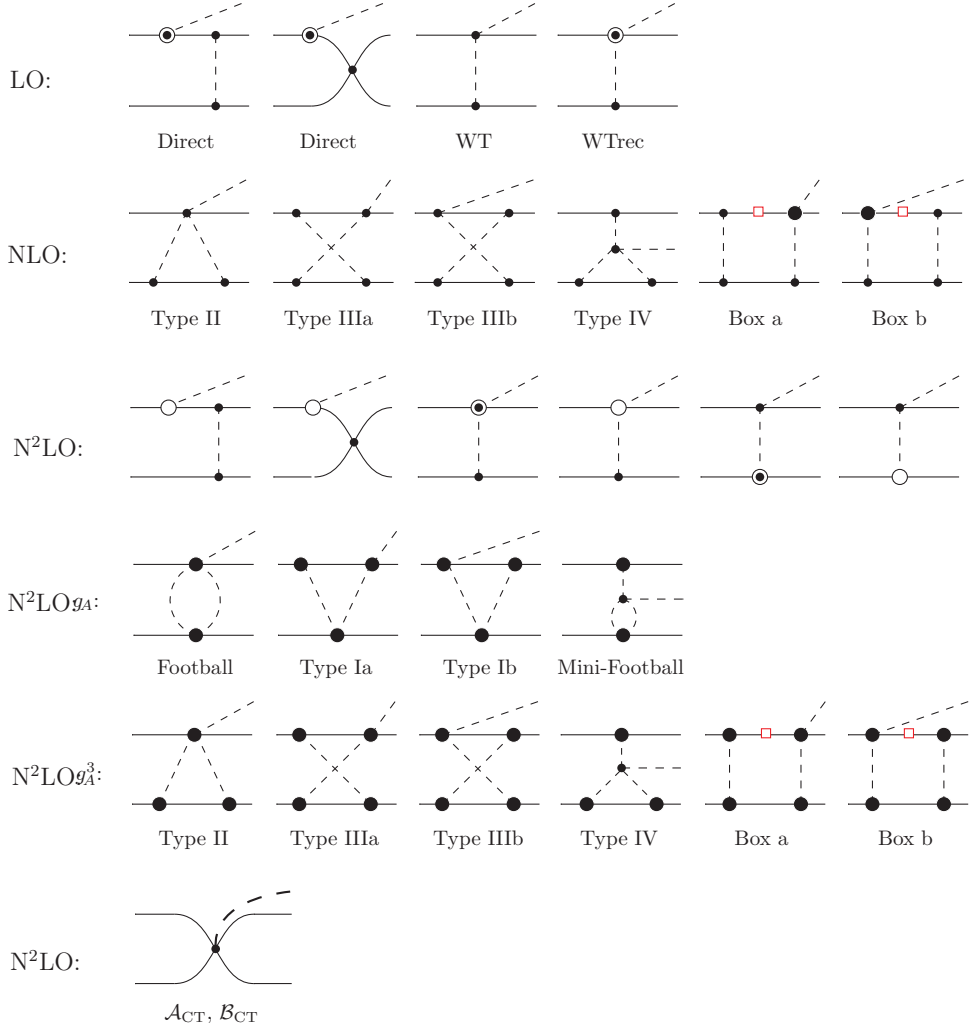}
\caption{\label{fig:allN2LO}
Complete set of diagrams up to N$^2$LO   (in the $\Delta$-less theory)  for  s-wave pions. 
Solid (dashed) lines denote nucleons (pions).
Solid dots correspond to the leading  vertices from
${\cal L}^{(1)}_{\pi\!N}$ and ${\cal L}_{\pi\pi}^{(2)} $,    $\odot$ stands for the sub-leading vertices from  ${\cal L}^{(2)}_{\pi\!N}$
whereas the blob  indicates the possibility to have  
both leading and subleading vertices from  ${\cal L}^{(1)}_{\pi\!N}$ and  
${\cal L}^{(2)}_{\pi\!N}$ (see  Fig. \ref{fig:typeII}  for  an  illustration),  
the opaque symbol $\circ$  stands for  the  
vertices $\sim1/m_N^2$ from  ${\cal L}^{(3)}_{\pi\!N}$.
The $NN$ contact interaction in the top row 
is represented by the leading S-wave LECs  $C_S$ and $C_T$ from ${\cal L}_{NN}$ whereas the 
contact five-point vertices in the bottom row 
are given by the LECs ${\cal A}_\text{CT}$ and ${\cal B}_\text{CT}$.
The red square in the box diagrams indicates 
that the corresponding nucleon propagator cancels with
parts of the $\pi N$ vertex and leads to the  irreducible contribution, 
see text for further details.
}
\end{figure}
%%%%%%%%%%%%%%%%%

The most general form of the threshold amplitude 
(where the pion is in an s-wave relative to a $NN$ S-wave final state) for
the pion-production reaction $N_1(\vec p\,)+N_2(-\vec p\,) \to N+N +\pi$ in
the center-of-mass frame 
can be written as \cite{NNLOswave}:
\begin{eqnarray}
M_{th}(NN\to NN \pi) =  \, {\cal A}\,(\vec \sigma_1 \times \vec \sigma_2)\cdot \vec p
\,\,\, \boldtau_+\cdot \boldphi^{\,*}  
+{\cal B}\,
\,(\vec \sigma_1 + \vec \sigma_2)\cdot \vec p  \,\,\, (-i)\, \boldtau_\times\cdot
\boldphi^{\,*} \, ,
\label{eq:Mthr}
\end{eqnarray}
where $\boldtau_+=\boldtau_1+\boldtau_2, \boldtau_\times=i \, \boldtau_1 \times \boldtau_2$  and 
$\vec\sigma_{1,2}$ and $\boldtau_{1,2}$ are the spin and isospin operators of
nucleons 1 and 2. This expression 
 incorporates  the  
%Pauli  
selection rules  for  the  $NN$ states. 
The final  pion's  isospin state
is denoted by $\boldphi$, e.g. $\boldphi= (0,0,1)$
for $\pi^0$-production and $\boldphi = (1,i,0)/\sqrt2$ for $\pi^+$-production.
For example, the   amplitude $\cal A$   corresponds  to  the production  of 
an s-wave pion accompanied  with the final state  
 spin-singlet  S-wave   $NN$  interaction  ($pp \to pp\pi^0$),  
 while  $\cal B$  corresponds  to  the  spin triplet $NN$ final state ($pp \to d\pi^+$).  

%%%%%%%%%%%%%%%%%%%%%%%%%%%%%%%%%%%%%%%%%%%%%%%%%%%%%%%%%%%%%%%%%%
\begin{figure}[t]
\includegraphics[width=14cm]{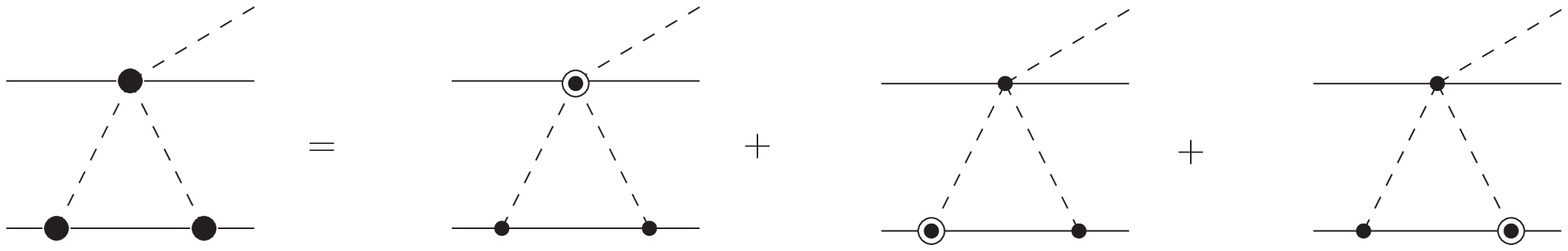}
\caption{\label{fig:typeII}
An  example  illustrating  the  notation   used  for  N$^2$LO  
operators  in  Fig.\ref{fig:allN2LO}. 
The sub-leading vertex should only appears at one vertex in each diagram.  
}
\end{figure}
%%%%%%%%%%%%%%%%%%%%%%%%%%%%%%%%%%%%%%%%%%%%%%%%%%%%%%%%%%%%%%%%% 

 As mentioned in the introduction, the reaction $NN\to NN\pi$ at
threshold involves
momenta of  ``intermediate range'' 
$p\approx \sqrt{\mpi \mN}$, Eq.~(\ref{expansionpapar}). 
For near threshold s-wave pion production,
the outgoing two-nucleon pair has a low relative three-momentum $p'$
and appears therefore  predominantly in S-wave.
We therefore assign  $p'$ an order $\mpi$, i.e,  
the expansion parameter in Eq.~(\ref{expansionpapar}) 
is augmented by $\chi_{\rm MCS}\simeq p'/p\simeq \mpi/p \simeq p/\mN$.

The calculations are based on the effective chiral Lagrangian  
given  explicitly  in  Ref.~\cite{NNLOswave},   
see  Refs.~\cite{OvK,ulfbible,Fettes,Fettes2}    for more  details.  
To \NNLO{},   
one  needs  to  keep  the  corrections  $\sim 1/m_N^2$,
  as will be  shown below.      
This  means  that  we need  to keep vertices   $\sim 1/m_N^2$  from  the  pion-nucleon  
  Lagrangian    ${\cal L}^{(3)}_{\pi\!N}$~\cite{Fettes,Fettes2}. 
In Appendix~\ref{lagrang} the  terms in the Lagrangian   relevant   for  our  study  are  listed explicitly.

The diagrams containing only pion and nucleon degrees of freedom
that contribute to the reaction $NN\to NN\pi$
up to N$^2$LO in the MCS expansion, are shown in Fig.~\ref{fig:allN2LO}.
The  first row  of  diagrams  represents  contributions  at  LO.  
 In the first row the first two diagrams  are sometimes called the ``direct'' 
one-nucleon diagrams or  impulse-approximation diagrams  in the literature,
whereas the other two graphs  are called rescattering diagrams.  
At  NLO  for  s-wave  pion  production   loop  diagrams start  to  contribute,  
as  shown  by the  second row of  graphs.  
As  will be  reviewed  below, these NLO  amplitudes   cancel  
completely~\cite{HanKai,towards,ksmk09,NNLOswave}.  
At  \NNLO{},  there are several  contributing tree-level diagrams  
which are  topologically   
similar  to those  at LO  but  with  sub-leading  vertices  from  
${\cal L}^{(2)}_{\pi\!N}$  and  even  ${\cal L}^{(3)}_{\pi\!N}$.
These diagrams are shown in the  third  row  in  Fig.~\ref{fig:allN2LO}.  
In  addition,   one needs   to  account  for  the  pion-nucleon  loops 
which, at   this  order,   can  be  combined  in  two  series of amplitudes, one  
proportional  to     $\gA^3$ with a topology like the NLO pion-nucleon loop diagrams 
 and    one proportional to $\gA$. 
These diagrams are given in rows four and five in    Fig.~\ref{fig:allN2LO}, respectively. 
 To the  order we  are  working,  it  suffices  to  include  the
  sub-leading vertex  from ${\cal L}^{(2)}_{\pi\!N}$  only once  
(but we have to consider all permutations of ${\cal L}^{(2)}_{\pi\!N}$ acting on a vertex) 
in the loops while retaining  the  other 
   vertices  at   leading  order as illustrated in   Fig.\ref{fig:typeII}.

For illustration 
of how the order of diagrams are 
estimated (or counted) in the MCS, 
we concentrate on the 
tree-level  rescattering  diagrams in  Fig.~\ref{fig:allN2LO} and will compare the LO and 
 the sub-leading diagrams designated as  N$^2$LO in Fig.~\ref{fig:allN2LO}. 
First we consider the rescattering diagram  in the first and third row where the 
$\pi NN$ vertex on the lower nucleon line originates from ${\cal L}_{\pi N}^{(1)}$. 
In Eq.~(\ref{treePC}) below, 
the $p/\fpi$  and  $1/p^2$  in front of the 
curly bracket stand for  the dimensional estimate  of  the 
$\pi NN$  vertex  and  the pion propagator, 
whereas the expressions after the curly bracket multiplied  by   $1/\fpi^2$ 
originate from the 
rescattering vertices  and correspond  to  the  $\pi N $ scattering vertices  from   
${\cal L}^{(1)}_{\pi\!N}$   and  ${\cal L}^{(2)}_{\pi\!N}$  [first  row in  Eq.~(\ref{treePC})], 
  $ {\cal L}^{(2)}_{\pi\!N}$  [second  row in  Eq.~(\ref{treePC})]  
and   ${\cal L}^{(3)}_{\pi\!N}$ [third row]: 
\begin{eqnarray}\label{treePC}
\text{ $M_{\rm rescat}$  $\propto\frac{p}{\fpi}\; \frac{1}{p^2}\;\frac{1}{\fpi^2}$}
\left\{ 
  \begin{array}{l l}
   \text{$\mpi; \quad   \,\left(\text{or}\ \ \frac{\dis p^2}{\dis m_N^{\phantom{a}}}\right )$} 
\\ \\
       \text{$\frac{\dis \mpi{\phantom{a}}}{\dis m_N^{\phantom{a}}}p; 
      \quad ( \text{or}\ \ \mpi p \, c_i)$} 
\\ \\
     \text{ $\frac{\dis p^3}{\dis m_N^2}$;  \quad   $ \left(\text{or}\ \ 
  \frac{\dis p^3 c_i}{\dis m_N^{\phantom{a}}}\right )$}, 
  \end{array} \quad = \quad\text{$\frac{\chi_{\rm MCS}}{\fpi^3}$}   
  \left\{ 
 \begin{array}{l l}
   \text{1}  & \quad \text{---  LO}\\ \\
   \text{$\chi_{\rm MCS}^2$} & \quad \text{---  N$^2$LO}\\ \\
     \text{$\chi_{\rm MCS}^2$}, & \quad \text{--- N$^2$LO,}
  \end{array}\right. 
  \right.
\end{eqnarray}
The  LO  operator  scales as  $\sim \chi_{\rm MCS}/\fpi^3$  for  s-wave  pions.  
As seen in the last  two rows  of  terms in Eq.\eqref{treePC} 
both  the $1/m_N$ corrections  from    ${\cal L}^{(2)}_{\pi\!N}$  
and the  $1/m_N^2$  corrections  from    ${\cal L}^{(3)}_{\pi\!N}$ 
need  to  be  taken into  account   at N$^2$LO. 
Meanwhile, the inclusion of a $1/m_N^2$  correction in the loops 
results in a  higher-order diagram 
and is ignored. 
Analogously, the estimated order of the   
     ``direct''  and rescattering   contributions  with    
sub-leading $\pi NN$ vertices  from   
    ${\cal L}^{(2)}_{\pi\!N}$ and  ${\cal L}^{(3)}_{\pi\!N}$  
yield the same results.  For  
example,  for  the   
         ``direct''  contribution  one  finds   
$ [ p \mpi^2/ (\fpi m_N^2) ]  [ 1/\mpi ]  [ 1/\fpi^2 ] =  \chi_{\rm MCS}^3/\fpi^3 $,  
 which  generates an amplitude at N$^2$LO.
   To  obtain  this  we  used that      the   first  term  on the l.h.s   
corresponds  to  the   $\pi NN$ vertex from ${\cal L}^{(3)}_{\pi\!N}$,   
the  next term  reflects the energy $v\cdot p\sim m_\pi$ of  
the  nucleon propagator  
while   the last  term   corresponds  to the  estimate  of  the  
one-pion exchange (OPE) or the  contact term.  
     Notice that while  the  tree-level  contributions  at  N$^2$LO  
with  the  sub-leading  vertices  from ${\cal L}^{(2)}_{\pi\!N}$ were  already 
        taken into  account  in the  literature,  
see  e.g.  Refs.~\cite{cohen,rocha},   the  corrections  stemming   from 
${\cal L}^{(3)}_{\pi\!N}$ are new  
 and  derived  here  for  the  first time.

Before  discussing  the results  at  \NNLO{}, one   comment  is  in order. 
We  noticed  above  that  the  loop  diagrams  at NLO  cancel  exactly.   
Here  
we  want  to  explain the cancellation pattern in more detail since  
it  is  quite general  and  
will be used  later   to establish  cancellations   
among the loops  at  N$^2$LO  both  with   nucleons  and Delta. 
 
\begin{figure}[t]
\includegraphics{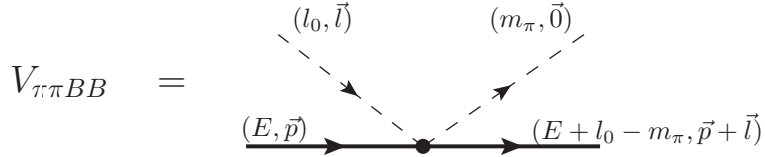}
\caption{\label{fig:vpipinn}
The pion-baryon ($\pi B\to \pi B$) transition vertex:
definition of kinematical variables as used in Eq.~(\ref{eq:pipivert}).  
Solid  thick lines  stand  for the  baryon (nucleon or $\Delta$) fields, 
dashed lines  denote  pions.}
\end{figure}

For the channel $pp\to pp\pi^0$, the sum of NLO diagrams of type II, III and IV in
Fig.~\ref{fig:allN2LO} is zero due to a cancellation between 
individual diagrams \cite{HanKai}, 
 while  the  box  diagrams  vanish due to  
the isovector  nature of the  WT  operator.
However,  the same diagrams II--IV give a finite contribution to the channel
$pp\to d\pi^+$ \cite{HanKai}.
As a result, the net contribution of these diagrams 
depends linearly on the $NN$
relative momentum which implies a large sensitivity to the
short-distance $NN$ wave functions~\cite{Gardestig}.
This seeming puzzle was solved in Ref.~\cite{towards}, where it was demonstrated
that for the deuteron channel there is an 
additional contribution at NLO, namely the
box diagrams in Fig.~\ref{fig:allN2LO}, 
stemming from the time-dependence of
the WT pion-nucleon  vertex.
To demonstrate this, we write the expression for the 
Weinberg--Tomozawa pion-baryon (nucleon or $\Delta$) scattering vertex
in the notation of Fig.~\ref{fig:vpipinn}
as:
%%%%%%%%%%%%%%%%
\begin{eqnarray}
V_{\pi\pi BB}&=&
l_0{+}\mpi {-}\frac{\vec l\cdot(2\vec p+\vec l)}{2m_B}
\nonumber \\
&=&
{2\mpi}+{\left(l_0{-}\mpi {+}E{-}\frac{(\vec l+\vec p)^2}{2m_B}-\delta +i\nolik \right)}-
{\left(E{-}\frac{\vec p\, ^2}{2m_B}-\delta +i\nolik\right)}
\ ,
\label{eq:pipivert}
\end{eqnarray}
%%%%%%%%%%%%%%%%%
where we kept the leading WT vertex and its baryon recoil correction,
which are of the same order in the MCS, as explained above. 
If the baryon line is a nucleon one has $\delta=0$, whereas 
$\delta \ne 0$ for the case of the $\Delta$.
%%%%%%%%%%%%%%%%%
%%%%%%%%%%%%%%%%
For simplicity, we omit the isospin dependence of the vertex.
The first term in the last line is the WT-vertex for  kinematics corresponding to the
on-shell incoming and outgoing baryons,
the second term is the
inverse of the outgoing baryon propagator while the 
third one is the inverse of the
incoming baryon propagator. 
Note that for on-shell
incoming and outgoing baryons, the expressions in brackets 
in \eqref{eq:pipivert} vanish, and
the $\pi$ baryon scattering vertex
takes its on-shell value $2\mpi$ 
(even if the incoming pion is off-shell).
A second consequence of Eq.~(\ref{eq:pipivert}) is that  only the
first term leads to a reducible diagram  when the rescattering diagram with the
$\pi B\to \pi B$  vertex
is convoluted with $NN$ wave functions.
The second and third terms in Eq.~(\ref{eq:pipivert}), however,
lead to irreducible contributions, since one of the baryon  
 propagators gets cancelled.  This cancellation is illustrated by red squares
on the nucleon propagators in the two box diagrams of Fig.~\ref{fig:allN2LO}. 
It was shown explicitly in Ref.~\cite{towards} that those
induced irreducible contributions
cancel exactly the finite remainder of the NLO loops (II--IV)
in the $pp\to d\pi^+$ channel.
As a consequence, there are no contributions at NLO  for both
$\pi^0$ and $\pi^+$ productions, see also our results in the two 
first rows of Tables~I  and II   of  Ref.~\cite{NNLOswave}.

In Ref.~\cite{NNLOswave} we extended  the analysis of the previous studies and
evaluated  the contribution  from pion-nucleon loops at \NNLO{}.  
As  follows from  Eq.\eqref{eq:pipivert},  
the  part  of the  $\pi N $ vertex   in the   diagrams  IIIa, IIIb  
and Box a, Box b  (the operators $\propto g_A^3$) 
 and     Ia  and Ib   (the operators $\propto g_A$)  cancels   
the  nucleon propagator  yielding  the  contribution   that  has  a  topology  
of   the  diagrams  II     or    ``football'',  respectively.  
Interestingly,  the  diagrams IV  and ``mini-football''   also  acquire    contributions  topologically  
similar  to   the  diagrams  II  and  ``football''  after canceling    one   of  the  pion  propagators
by a part of the four-pion vertex.  
 In conclusion,   Ref.~\cite{NNLOswave}  found that    
only those parts of the $g_A^3$ ($g_A$)-diagrams 
which cannot be reduced  to the topology of the  diagram II  (football) 
in Fig.~\ref{fig:allN2LO},
give a non-zero contribution to the transition amplitude. 
Thus,  only very few \NNLO{} contributions to the pion-production amplitude remain.   
Especially, all recoil  corrections  $\propto 1/m_N$  
and also  all  corrections  $\propto c_i$  vanish  
completely in the sum of the loop diagrams at \NNLO{}. 
The non-vanishing amplitudes emerge from the  diagrams  IIIa,  IIIb 
and  IV  in  the  case  of the $g_A^3$-graphs  
and  from  the  diagrams  Ia,  Ib  and  ``mini-football'' for  the  $g_A$-operators, 
where  all the vertices,  except  for the  recoil correction to the  WT  vertex,  
originate from  the leading Lagrangian    ${\cal L}^{(1)}_{\pi\!N}$.
The explicit expressions of the resulting amplitudes are given in Sec.\ref{sec:loops}.

%%%%%%%%%%%%%%%%%%%%%%%%%%%%%%%%%%%%%%%%%%%%%%%%%%%%%%%%%%%%%%%%%%%%%%%%%%%%%%
\subsection{Pion-production operator   from tree-level  diagrams}
\label{sec:diagrams}
%%%%%%%%%%%%%%%%%%%%%%%%%%%%%%%%%%%%%%%%%%%%%%%%%%%%%%%%%%%%%%%%%%%%%%%%%%%%%%%
%%%%%%%%%%%%%%%%%
\begin{figure}[h] 
\includegraphics[width=16.5cm]{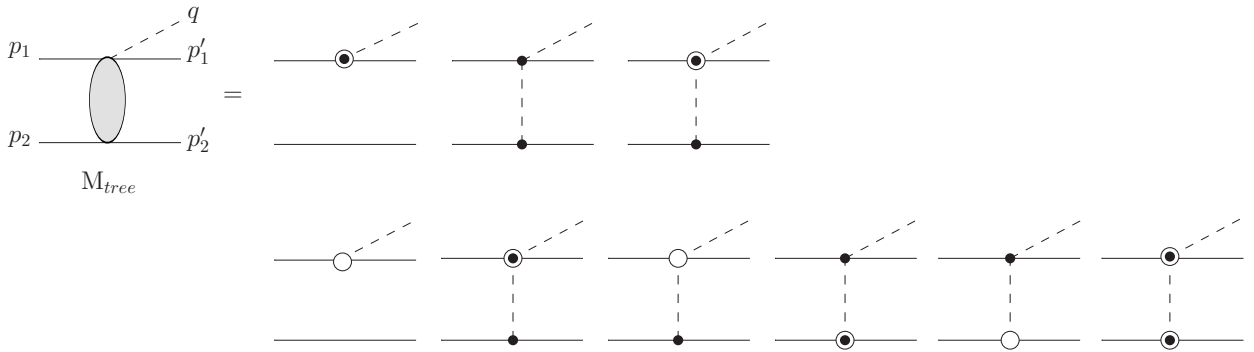}
\caption{ \label{fig:treeN2LO}
Single-nucleon and rescattering diagram  contributions  to  s-wave pion production up to \NNLO{}:   
the diagrams in the  first  row  on  the r.h.s  appear  at LO,   
the   diagrams in the second row  give the \NNLO{}  contributions.     
The last diagrams  in  the first and second  rows  involve the WT recoil correction,
whereas the second diagram in the second row involves 
the $c_i$ rescattering vertex.  
}
\end{figure}
%%%%%%%%%%%%%%%%%%%%
 
In this  section  we  derive   the   amplitudes from  diagrams 
shown  in Fig.~\ref{fig:treeN2LO}  
up-to-and-including  \NNLO{}  in the theory 
involving pion  and  nucleon degrees of  freedom for  s-wave  pion  production.   
 The  tree-level and  loop  contributions  due to  the explicit  Delta-resonance  
will  be  discussed in Secs.~\ref{sec:Deltatree} and \ref{sec:Deltaloops}.

The rescattering operator  at  LO  involves  the Weinberg--Tomozawa $\pi N$ vertex  from    
${\cal L}^{(1)}_{\pi\!N}  $   and   its   recoil correction  from   
${\cal L}^{(2)}_{\pi\!N}  $ which give the resulting operator amplitude:
\begin{eqnarray}
	i M^\text{LO}_{\rm rescat} = i M_\text{WT}^\text{LO}+i M_\text{WT}^\text{recoil}
	=\frac{g_A}{2 f_\pi^3} 
	\frac{ v \cdot q}{k_2^2-m_\pi^2+i0}
		(S_2 \cdot k_2) \taux^a \, + (1 \leftrightarrow 2),
\label{eq:treeLO}
\end{eqnarray}
where the superscript $a$ ($a$=1,2,3)  here  and in what follows  
refers to the isospin quantum number of  the outgoing pion field.   
The  momenta are defined  in Fig.\ref{fig:treeN2LO}. Further,   
$k_2=p_2-p_2'$,   $v^{\mu} = (1,\vec{0})$ 
is the nucleon four-velocity 
and $S^{\mu}$ 
is its spin-vector,   see  Appendix  \ref{lagrang}  for  further details.

 The  rescattering operator  at  \NNLO{}  contains   the  corrections 
suppressed as  $1/m_N$  due to  the vertices  from ${\cal L}^{(2)}_{\pi\!N}  $   and 
also the corrections  $\propto 1/m_N^2$  from ${\cal L}^{(3)}_{\pi\!N}$. 
We  call these amplitude operators 
 $M_{\rm rescat1}^\text{N$^2$LO}$ and  $M_{\rm rescat2}^\text{N$^2$LO}$, respectively.  
The  explicit expressions are:
%%%%%%%%%%%%%%%%
\begin{eqnarray}
	i M_{\rm rescat1}^\text{N$^2$LO} &=& \frac{g_A}{f_\pi^3} \frac{ (S_2 \cdot k_2) 
                 \tau_2^a }{k_2^2-m_\pi^2+i0}
		\left[  
			4c_1 m_\pi^2 - v \cdot q \, v \cdot k_2 
                         \left( 2c_2+2c_3-\frac{g_A^2}{4 m_N} \right) 	
		\right]  
\nonumber 
\\ 
            &&- \frac{g_A}{f_\pi^3}
	\frac{(v\cdot q )\,  \taux^a }{k_2^2-m_\pi^2+i0}
           \frac{S_2 \cdot (p_2+p_2^\prime)}{4 m_N}  (v \cdot k_2 )
            +   (1 \leftrightarrow 2),  
\label{treeNNLO1} 
\\
	i M_{\rm rescat2}^\text{N$^2$LO} &=& 
	\frac{g_A}{f_\pi^3}
	\frac{v\cdot q }{k_2^2-m_\pi^2+i0} 
	\Bigg\{ 
	- \tau_2^a \, (S_2 \cdot k_2) 
	  \frac{k_2 \cdot (p_1+p_1^\prime)}{m_N^2} \left( m_N c_2 -\frac{g_A^2}{16} \right) 
\nonumber
\\ 
	&&+ \taux^a \, (S_2 \cdot k_2) 
		\left[
			\frac{\vec{p}_1^{\,2}+\vec{p}_1^{\,\prime 2}}{16 m_N^2}
			+ \frac{1+g_A^2+8m_N c_4}{8 m_N^2} 
                 \left( [S_1\cdot k_2, S_1 \cdot (p_1+p_1^\prime)] + \frac{k_2^2}{2}\right) 
		\right]
\nonumber
\\	&&- \frac{\taux^a }{8 m_N^2} 
		\left[  
			(S_2 \cdot p_2^\prime) p_2^2 -  (S_2 \cdot p_2) p_2^{\prime 2}
		\right] 
		\Bigg\} +  (1 \leftrightarrow 2),
\label{treeNNLO2}
\end{eqnarray}
%%%%%%%%%%%%%%
%
where $\left[ S_{1\mu}, \, S_{1\nu}\right] =S_{1\mu} S_{1\nu} - S_{1\nu} S_{1\mu}$.
The  first  two  terms  in the curly bracket  in Eq.\eqref{treeNNLO2}   
are  due to the corrections  to  the $\pi \pi N  N$  vertex from ${\cal L}^{(3)}_{\pi\!N}$ 
while  the  last   one stands  for  the   correction  to the  
$\pi NN$ vertex  at the  same order.  
Both  amplitudes $M_{\rm rescat1}^\text{N$^2$LO}$  and $M_{\rm rescat2}^\text{N$^2$LO}$ contribute  to  the  
isoscalar  ($\cal{A}$) and   isovector  ($\cal{B}$) amplitudes.

In  addition to the rescattering operators just derived,  
one  needs  to   account  for  the  contributions from the  direct pion  
emission from  a single nucleon,  the so-called  direct  diagrams  
which are  shown  in  
the first and third rows of Fig.\ref{fig:allN2LO} and contribute  at  LO  and \NNLO{}.  
Note     
that  in these direct diagrams, 
the  OPE  or the $NN$ contact term, 
which  appear together  with the $\pi NN$ vertex  for outgoing pion  in Fig.\ref{fig:allN2LO},  
will  be  considered  as   part 
of the final (or initial)  $NN$  wave  function.  
In Fig.~\ref{fig:treeN2LO}, the OPE and the $NN$ contact term are, therefore, not shown.  
After this separation of the  $NN$-interaction part,  
the  contribution  of   the  ``direct''  diagrams 
shown in  Fig.\ref{fig:treeN2LO}
becomes a one-nucleon operator and can be  written as  
%%%%%%%%%%%%%%% 
\begin{eqnarray}
	i M_{\rm dir}& =& \frac{g_A}{f_\pi} \tau_1^a  \, v \cdot q \ 
            \delta(\vec p_2- {{\vec p}_2}{'}) 
\nonumber 
\\
&&\times
	\left[ 	- \frac{1}{2 m_N}  S_1 \cdot (p_1+p_1^\prime)
		+ \frac{1}{4 m_N^2} \left( v \cdot p_1 (S_1 \cdot p_1) 
		+ v \cdot p_1^\prime (S_1 \cdot p_1^\prime)\right) 
	\right] +   (1 \leftrightarrow 2)
\label{Mdir}
\end{eqnarray} 
%%%%%%%%%%%%%
This  amplitude   contributes to observables  only  when  convoluted  with the 
$NN$ wave  functions.
For further discussion see Appendix \ref{convol}.

These LO and \NNLO{} operator amplitudes generated by the 
diagrams in Fig.\ref{fig:treeN2LO} have to be added 
to the \NNLO{} loop diagrams to be presented 
in Sec.~\ref{sec:loops}  
in order to generate the 
complete \NNLO{} s-wave pion-production operator 
amplitude from pion-nucleon diagrams. 
We postpone the discussion of the combined s-wave pion-production amplitude 
until the end of Sec.~\ref{sec:swaveD} where also the $\Delta$
degrees of freedom will be added to the pion production  s-wave amplitude.

%%%%%%%%%%%%%%%%%%%%%%%%%%%%%%%%%%%%%%%%%%%%%%%%%%%%%%%%%%%%%%%%%%%%%%%%%%%%%%%%%%%%%%%%%
\subsection{Pion-production  operator   from pion-nucleon loop  diagrams  up  to  \NNLO{} }
\label{sec:loops}
%%%%%%%%%%%%%%%%%%%%%%%%%%%%%%%%%%%%%%%%%%%%%%%%%%%%%%%%%%%%%%%%%%%%%%%%%%%%%%%%%%%%%%%%

The  contribution  of  pion-nucleon  loops  to  the  production  operator   for  
s-wave  pions  was  derived in   Ref. \cite{NNLOswave} and we just summarize the results:   
\begin{eqnarray}
i M^{\text{\NNLO{}}}_{\gA} &=& \frac{\gA \ (v \cdot q)}{\fpi^5}
\taux^a  (S_1 + S_2)\cdot k_1
\left[ \frac16 I_{\pi\pi}(k_1^2) - \frac{1}{18} \frac{1}{(4 \pi)^2} \right] \; ,
\label{eq:Mga1final}
\\
    i M^{\text{\NNLO{}}}_{\gA^3} &=& \frac{\gA^3 \ (v \cdot q)}{\fpi^5} \bigg\{
    \taup^a \, i \varepsilon^{\alpha\mu\nu\beta} v_\alpha k_{1\mu} S_{1\nu} S_{2\beta}
    \left[-2 I_{\pi\pi}(k_1^2)\right] 
\nonumber 
\\
    && + \taux^a \,  (S_1 + S_2) \cdot k_1
    \left[ -\frac{19}{24}  I_{\pi\pi}(k_1^2) + \frac{5}{9} \frac{1}{(4 \pi)^2} \right]
    \bigg\}, 
\label{eq:Mga3final}
\end{eqnarray}
where  the integral  $\ipipi(k_1^2)$  is  defined  in  
Appendix~\ref{sec:appbasicint}\footnote{
%%%%%%%%%%
In Ref. \cite{NNLOswave},  
the  integral $\ipipi(k_1^2)$  was  called  $J(k_1^2)$.  
%%%%%%%%%
}.
Note that both $M^{\text{\NNLO{}}}_{\gA}$ and $M^{\text{\NNLO{}}}_{\gA^3}$
are proportional to the outgoing pion energy $v\cdot q \simeq \mpi$, i.e.
both operator amplitudes vanish at threshold in the chiral limit as expected.

The contributions of the loops to the amplitudes
${\cal A}$ and ${\cal B}$, see Eq.~\eqref{eq:Mthr},
can be separated into  singular and  finite parts,
where the singular parts are given by the  
 UV divergences appearing in the integral, $\ipipi(k_1^2)$,  
in Eqs.~(\ref{eq:Mga1final}) and (\ref{eq:Mga3final}).  
\begin{eqnarray}
{\cal A}&=& \frac{\mpi}{(4\pi \fpi)^2\fpi^3}(\tilde {\cal A}_\text{singular}
+\tilde{\cal A}_\text{finite} ), \nonumber \\
{\cal B}&=&\frac{\mpi}{(4\pi \fpi)^2\fpi^3} (\tilde {\cal B}_\text{singular}
+\tilde {\cal B}_\text{finite} ).  
\label{eq:loops1}
\end{eqnarray}
The UV divergences are 
absorbed into  LECs accompanying the $(NN)^2\pi$ amplitudes 
${\cal A}_\text{CT}$ and ${\cal B}_\text{CT}$,  
given in the last row  in  Fig.\ref{fig:allN2LO}. 
By renormalization,  the singular parts of the loop amplitudes 
are eliminated and we 
are left with the renormalized finite LECs, 
${\cal A}^r_\text{CT}$ and ${\cal B}^r_\text{CT}$, which will be added 
to the finite parts of the loop amplitude operators. 
Based on the renormalization scheme of Ref.~\cite{NNLOswave}, 
the finite parts of the pion-nucleon loops  are 
\begin{eqnarray}
\label{ABfin}
\tilde {\cal A}_\text{finite}(\mu)&=& -\frac{g_A^3}{2}
\left[ 1 - \log \left( \frac{\mpi^2}{\mu^2} \right)
-  2 F_1 \left(\frac{-{\vec p\,}^2}{\mpi^2}\right) \right],
\nonumber \\
 \tilde {\cal B}_\text{finite}(\mu)&=& -\frac{g_A}{6}
\left[ -\frac{1}{2}\left(\frac{ 19}{4} g_A^2 -1\right)
\left( 1 - \log \left( \frac{\mpi^2}{\mu^2} \right)
-  2 F_1 \left(\frac{-{\vec p\,}^2}{\mpi^2}\right) \right)+ 
    \frac{5}{3}g_A^2-\frac16
 \right] \ , 
\end{eqnarray}
where  $\mu$ is the scale and 
the  function $F_1$  is  defined  in  the appendix,  Eq.~\eqref{eq:F1}. 
In general, as discussed in Ref.~\cite{NNLOswave}, 
the finite parts of the loops 
$ \tilde {\cal A}_\text{finite}$ and  $ \tilde {\cal B}_\text{finite}$
can be further decomposed into the {\it short}- and {\it long}-range parts.
The former one is just a (renormalization scheme dependent) constant 
to which  all terms in Eq.~\eqref{ABfin} except $F_1$ contribute.
On the other hand, the long-range part of the loops is scheme-independent. 
By expanding the function $F_1(-{\vec p\,}^2/m_\pi^2)$, Eq.~(\ref{eq:F1}), 
which is the only long-range piece in 
Eq.~\eqref{ABfin}, in the kinematical
regime relevant for pion production, i.e.~$({\vec p\,}^2/m_\pi^2) \gg 1$, 
one obtains 
up-to-and-including
\NNLO{} terms 
\begin{eqnarray}
\tilde {\cal A}^\text{long}_\text{finite}&=&-\frac{g_A^3}{2} 
\log\left(\frac{\mpi^2}{{\vec p\,}^2}\right) + 
{\cal O}\left(\frac{\mpi^2}{{\vec p\,}^2}\right),
\nonumber \\
 \tilde {\cal B}^\text{long}_\text{finite}&=& \frac{g_A}{12}
\left(\frac{ 19}{4} g_A^2 -1\right)
\log\left(\frac{\mpi^2}{{\vec p\,}^2}\right)+ 
{\cal O}\left(\frac{\mpi^2}{{\vec p\,}^2}\right).
\end{eqnarray}

A numerical evaluation of these terms     gives 
$ \tilde {\cal A}^\text{long}_\text{finite}=2.2$ and 
$ \tilde {\cal B}^\text{long}_\text{finite}=-1.5$.  
In Ref.~\cite{NNLOswave} these  numbers were  compared    
with  those from  the most important  phenomenological contributions which were proposed in   
Refs.~\cite{Lee,HGM,Hpipl,jounicomment}  in order to resolve the discrepancy 
between phenomenological 
calculations and experimental data.    
Using the mechanisms suggested in Refs.~\cite{Lee,HGM,Hpipl,jounicomment},  one 
obtained 
$\tilde{\cal A}^r_\text{CT}\simeq 2$ and $\tilde{\cal B}^r_\text{CT}\simeq 1$ 
in the  same units.   
Based  on this,  it   was  concluded in \cite{NNLOswave} 
 that  the  scheme-independent  long-range  contributions 
  of  pion-nucleon loops,  not included  in the previous studies,  
    are  comparable in size  with the  contribution  
needed to  bring    theory in agreement  with    experiment.

In the next section, Sec.\ref{sec:swaveD}, we derive  the  results    for \NNLO{} loops 
including the Delta resonance.  In particular,  it  will  be  shown  
that the long-range  contributions  of  the  pion-nucleon  and the Delta 
loops  are  of  similar  size,  in  agreement  with the power  counting  estimate.

%%%%%%%%%%%%%%%%%%%%%%%%%%%%%%%%%%%%%%%%%%%%%%%%%%%%%%%%%%%%%%%%%%%%%%%%%%%%%%%%%%%%%%%%%

\section{S-wave pion production to  \NNLO{}:    $\Delta$-resonance  induced contributions }
\label{sec:swaveD}
%%%%%%%%%%%%%%%%%%%%%%%%%%%%%%%%%%%%%%%%%%%%%%%%%%%%%%%%%%%%%%%%%%%%%%%%%%%%%%%%%%%%%%%%

The threshold pion-production reaction involves energies where the $\Delta$-resonance 
is not heavy enough to be parametrized just by $\pi N$ LECs. 
The $\Delta$ should in fact 
be explicitly included 
in the loops as virtual nucleon excitations in order for the effective theory 
to properly describe the physics in this energy region. 
Indeed,  whereas the mass difference $\delta$ 
is non-zero even in the chiral limit of the theory 
 (when $m_\pi \to 0$),    the physical value  of
$\delta$,  $\delta\approx  $ 300 MeV,   
is numerically  very close   to  the
``small''  scale   in the MCS,  i.e.  
the  momentum  scale $p \sim \sqrt{m_\pi  m_N}$.   
Hence,    Hanhart and Kaiser~\cite{HanKai} argued that, 
as a practical ``consistency'' in MCS,     
$\delta$ should  be counted as   $p$. 
In this section  we will outline the operator structure   
due to the inclusion of 
explicit $\Delta$ degrees of freedom for the $NN \to NN \pi$ reaction.

When the Delta is explicitly included,  the LECs  $c_2$, $c_3$ and $c_4$,  
which are determined from pion-nucleon data,  
have to be re-evaluated.  As a  consequence,  
in the  Delta-full theory, one obtains  the  LECs  in which  the  
Delta contribution is  subtracted. 
These  residual     LECs  enter  the calculation of  the  
pion-production  operator  derived in Sec. \ref{sec:nucl}, see 
Eqs.\eqref{treeNNLO1} and \eqref{treeNNLO2}.    
Once we include explicitly the  $\Delta$-field 
in the Lagrangian,  the 
parameter  $\delta \sim p$ will appear in loops containing the $\Delta$ propagator 
and the  resulting  loop momentum will naturally be
of the order  of  $p$,  i.e.,  
these loop diagrams will then contribute at NLO and \NNLO{} in the MCS.

In the MCS with a $\Delta$ explicitly included,
we also have to consider   
loop diagrams with topologies different from those
discussed in previous sections. 
Some of these additional loop diagrams containing a $\Delta$ 
propagator will renormalize 
LO diagram vertices as well as 
the nucleon wave function. 
This is in contrast to  the loop diagrams with only nucleon and pion propagators,  which
contribute  to  renormalization of the vertices at \NNNNLO{}  only,  
as  argued in Ref.~\cite{NNLOswave}.  
This is a higher order 
than what is considered in this work. 
These considerations will be explained in detail in the next subsections.

%%%%%%%%%%%%%%%%%%%%%%%%%%%%%%%%%%%%%%%%%%%%%%%%%%%%%%%%%%%%%%%%%%%%%%%%%%%%%%%
\subsection{The Lagrangian with $\Delta$ interactions}
\label{sec:DeltaLag}
%%%%%%%%%%%%%%%%%%%%%%%%%%%%%%%%%%%%%%%%%%%%%%%%%%%%%%%%%%%%%%%%%%%%%%%%%%%%%%

The  evaluation of   the amplitude contributions involving $\Delta$ 
is based on  the  effective Lagrangian \cite{Hemmert,Hemmert:1997ye}
which  reads  in  
the $\sigma$-gauge 
\begin{eqnarray}
{\cal L}_{\pi N \Delta} &=& 
-\Psi_\Delta^\dagger (i v\cdot\partial - \delta)
\Psi_\Delta
 + \frac{g_1}{\fpi}  \Psi_\Delta^\dagger \, {\Sdd}^\mu S^\beta \Sd_\mu \, T_i \boldtau 
\cdot \partial_\beta \boldpi  T_i 
\Psi_\Delta
\nonumber
\\ 
&&- \frac{1}{4 f_{\pi}^{2}} 
\Psi_\Delta^\dagger \Big[ 
         (\dot{\boldpi}\times{\boldpi})  \cdot  {T}_i^\dagger \boldtau  { T}_i + 2i
         \left(   ({\bf T}^\dagger  \cdot\boldpi)  ({\bf T} \cdot
           \dot{\boldpi}) -
( {\bf T}^\dagger \cdot \dot{\boldpi})  ({\bf T} \cdot  \boldpi)
\right)     \Big]   \Psi_\Delta
\nonumber
\\ 
&&- \frac{h_A}{2f_\pi} \Big[ 
N^\dagger {\bf T}\cdot \left(\partial^\mu \boldpi
{+}\frac{1}{2f_\pi^2}\boldpi (\boldpi \cdot \partial^\mu \boldpi) \right)
      \Sd_\mu\, \Psi_\Delta 
+ h.c. \Big] 
\nonumber
\\
&&+
\frac{h_A}{2m_N f_\pi} \Big[ i
N^\dagger {\bf T}\cdot \dot{\boldpi}\
    \Sd \cdot \partial \Psi_\Delta 
+ h.c. \Big]+\cdots \, ,
\end{eqnarray} 
where   
$g_1$  is the  $\pi \Delta\Delta$  
coupling  constant, $g_1= 9/5 g_A$  from Ref.\cite{FettesDel}, 
$h_A$ is the leading $\pi N \Delta $ coupling constant and $\Sd$ and ${\bf T}$ 
are the spin and isospin transition matrices, 
normalized such that 
\begin{eqnarray}
\Sd_\mu \Sd_\nu^\dagger &=& g_{\mu \nu} - v_\mu v_\nu - \frac{4}{1-d} S_\mu S_\nu,	
\quad
T_i T_j^\dagger = \frac{1}{3}\left( 2\delta_{ij}-i\epsilon_{ijk}\, \tau_k\right) 
\, ,  \quad i,j=1,2,3 .
\nonumber 
\end{eqnarray}
An estimate of the $\pi N \Delta $ coupling
constant  $h_A= 2g_{\pi N\Delta} = 3g_A/\sqrt{2} =2.7$ is
derived from large $N_c$ arguments~\cite{Kaiser1998},
whereas a  determination of $g_{\pi N\Delta}$  from a fit  to the decay width of the
$\Delta$ resonance to leading order in the small-scale expansion
 gives $h_A=2.1$~\cite{Hemmert:1997ye}.

%%%%%%%%%%%%%%%%%%%%%%%%%%%%%%%%%%%%%%%%%%%%%%%%%%%%%%%%%%%%%%%%%%%%%%%%%%%%%%%
\subsection{Reducible diagrams  with  $\Delta$-resonance}
\label{sec:Delta_reduce}
%%%%%%%%%%%%%%%%%%%%%%%%%%%%%%%%%%%%%%%%%%%%%%%%%%%%%%%%%%%%%%%%%%%%%%%%%%%%%%

%%%%%%%%%%%%%%%%%%%%%%%%%%%%%%%%%%%%%%%%%%%%%%%%%%%%%%%%%%%%%%%%%%
\begin{figure}[t]
\includegraphics[scale=0.6]{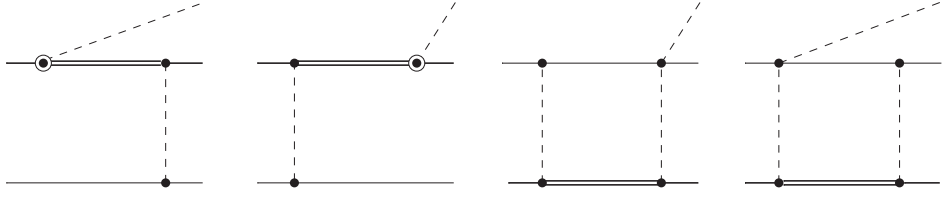} 
\caption{\label{fig:reduceD} 
Reducible  $\Delta$ contributions  at  NLO  (first two diagrams) and at \NNLO{}  (last  two).} 
\end{figure}
%%%%%%%%%%%%%%%%%%%%%%%%%%%%%%%%%%%%%%%%%%%%%%%%%%%%%%%%%%%%%%%%% 

%%%%%%%%%%%%%%%%%%%%%%%%%%%%%%%%%%%%%%%%%%%%%%%%%%%%%%%%%%%%%%%%%%
\begin{figure}[t]
\includegraphics[scale=0.6]{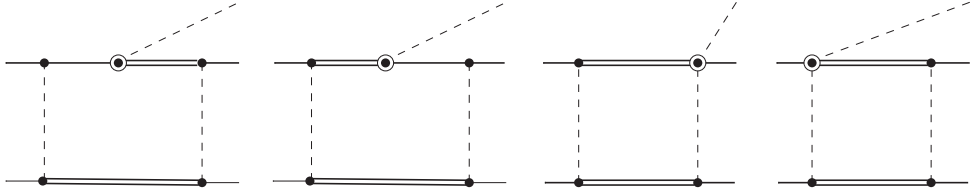} 
\caption{\label{fig:DD} 
Examples of N$^3$LO  contributions  involving  $\Delta \Delta$ intermediate  states.} 
\end{figure}
%%%%%%%%%%%%%%%%%%%%%%%%%%%%%%%%%%%%%%%%%%%%%%%%%%%%%%%%%%%%%%%%% 

The  application  of the         
  scheme  originally proposed  by Weinberg~\cite{wein90,weinberg1991}  
to pion-production  reactions suggests  that   the    
  full pion-production amplitude  can be  evaluated  by   convolving  
the  pion-production operator,  which   consists of  irreducible 
  graphs only,   with $NN$  wave  functions  in the  initial  and  final  states.   
Only on-shell amplitudes are physically meaningful and 
can be compared to each other  via  power  counting. 
To find the MCS order we estimate the order of  the  diagrams shown in
Fig. \ref{fig:reduceD}.   It is understood that       
the OPE in the diagrams will in the actual evaluations be considered as a part of  the  
$NN\to N\Delta$  amplitude 
similar to the direct diagrams discussion in Sec.~\ref{sec:diagrams}.  
For example,   the first  diagram in Fig. \ref{fig:reduceD}  is  
suppressed  relative  to  the  LO direct  one-nucleon diagram in Fig.~\ref{fig:allN2LO} 
since the  $\Delta$ propagator is  reduced  by  $\mpi /\delta\sim \mpi  /p$ 
compared to the nucleon one 
given in  Eq.~(\ref{treePC}).  
Therefore,  this direct $\Delta$-diagram in Fig.~\ref{fig:reduceD} contributes  at NLO.  
Meanwhile,  the reducible  
box  diagrams  
(the last  two) in Fig.~\ref{fig:reduceD},  which  are to  be   evaluated  with  
the  on-shell  (2$\mpi$)  $\pi N $ vertex   as  shown in  Eq.\eqref{eq:pipivert},
start  to  contribute  at N$^2$LO.   
Interestingly,  the  diagrams  in  Fig.~\ref{fig:DD}  involving  the $\Delta \Delta$ intermediate  state   contribute to the  
on-shell production operator  only  with subleading vertices at
 the loop-level   and  are therefore  of a higher order.  
In particular,  a comparison  of the  
 box  diagrams  with  one and  two  $\Delta$
 propagators  in Figs. \ref{fig:reduceD} and \ref{fig:DD} reveals  
that  the latter  are  suppressed  due  to  the  absence  of  the  
$\pi N\to \pi \Delta$ vertex  at  leading order.  

%%%%%%%%%%%%%%%%%%%%%%%%%%%%%%%%%%%%%%%%%%%%%%%%%%%%%%%%%%%%%%%%%%
\begin{figure}[t]
\includegraphics[scale=0.6]{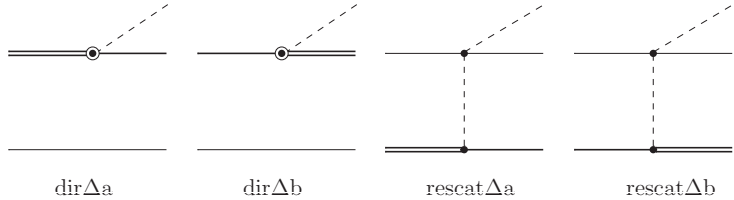}
\caption{\label{fig:treeDelta}  
Single baryon and rescattering diagrams with 
$\Delta$ contributions which appear  as  building blocks  in the  construction of 
the pion-production amplitude.  In the last  two rescattering 
diagrams only  the   on-shell  part  of   the  $\pi N$  vertex  
(2\mpi)  from  Eq.~\eqref{eq:pipivert}  
should  be  included.}
\end{figure}
%%%%%%%%%%%%%%%%%%%%%%%%%%%%%%%%%%%%%%%%%%%%%%%%%%%%%%%%%%%%%%%%% 

Since  the $NN$  and $N\Delta$  states 
are coupled  in the  $NN$ models \cite{CCF,CDBonn} which will be  used  to  
evaluate  the  initial  and  final state $NN$  
wave  functions in $NN\to NN\pi$,  
the  full pion-production amplitude  also receives  contributions   from  
the  building  blocks  containing  $N\Delta$  states as    shown in 
Fig.\ref{fig:treeDelta}.   
In  full  analogy  to the  ``direct''  single nucleon diagrams in Fig.~\ref{fig:allN2LO} 
and   as   discussed  in Sec.~\ref{sec:nucl},    diagrams 
shown  in  Fig. \ref{fig:treeDelta}    do  not  contribute  to  
the on-shell  pion-production operator  but  are relevant  only  when convolved 
with the  $NN-N\Delta$  amplitude  either  in the  initial  
or in the  final state.  
%%%%%%%%%%%%%%%%%%%%%%%%%

%%%%%%%%%%%%%%%%%%%%%%%%%%%%%%%%%%%%%%%%%%%%%%%%%%%%%%%%%%%%%%%%%%%%%%%%%%%%%%%
\subsection{Tree-level  diagrams with $\Delta$-resonance}
\label{sec:Deltatree}
%%%%%%%%%%%%%%%%%%%%%%%%%%%%%%%%%%%%%%%%%%%%%%%%%%%%%%%%%%%%%%%%%%%%%%%%%%%%%%

In this  subsection we provide  explicit  expressions for the  
 contributions  of  the  diagrams in Fig.\ref{fig:treeDelta}, 
and we obtain the following expressions: 
\begin{eqnarray}
i M_{\rm dir\Delta a}  &=& 
    -\frac{\gpind}{\mN \fpi} T_1^a \vdotq (\Sd_1 \cdot p_1)     \delta(\vec p_2- {{\vec p}_2}^{\,\prime}),
    \nonumber 
\\
i M_{\rm dir\Delta b}  &=&	
    -\frac{\gpind}{\mN \fpi} T_1^{\dagger a} \vdotq (\Sdd_1 \cdot p_1^\prime)     \delta(\vec p_2- {{\vec p}_2}^{\,\prime}),
    \nonumber
\\
i M_{\rm rescat\Delta a}&=& 
    \frac{\gpind}{2 \fpi^3} \vdotq \, i \epsilon^{bac} \tau_1^c T_2^b 	
    \frac{1}{k_2^2-\mpi^2 + i0} (\Sd_2 \cdot k_2),
    \nonumber
\\
i M_{\rm rescat\Delta b}&=&	
    \frac{\gpind}{2 \fpi^3} \vdotq \, i \epsilon^{bac} \tau_1^c T_2^{\dagger b} 	
    \frac{1}{k_2^2-\mpi^2 + i0} (\Sdd_2 \cdot k_2).
    \label{Mdel}
\end{eqnarray}
Of course, these  tree-level  pion-production amplitudes with a $N\Delta$ initial or final
state can  not  by definition  contribute  to the  $NN\to NN \pi$ 
irreducible production operator.  
On the other hand,  the amplitudes \eqref{Mdel} give a nonzero contribution 
to the full pion-production amplitude
  when they are inserted as a building block into
those of FSI and ISI diagrams which have an $N\Delta$  intermediate state.   
The corresponding expressions  are  given  in  the  Appendix  \eqref{convol},    
see  Eqs.  \eqref{MFSI} \eqref{MISI} and  \eqref{MISIFSI}.
The contribution of these  operators  corresponding to   charged 
pion production in $pp\to d\pi^+$  was    
evaluated in Ref.~\cite{NNpiMenu}.

%%%%%%%%%%%%%%%%%%%%%%%%%%%%%%%%%%%%%%%%%%%%%%%%%%%%%%%%%%%%%%%%%%%%%%%%%%%%%%% 
\subsection{Loop diagrams with $\Delta$ propagators}
\label{sec:Deltaloops}
\begin{figure}[t]
\includegraphics[scale=0.6]{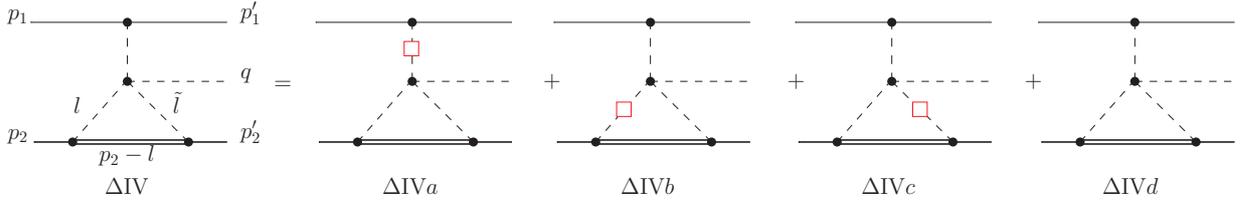}
\caption{\label{fig:DIV}
An example  of the  loop  diagrams  with  the explicit $\Delta$. 
Double lines denote the $\Delta$ propagator,  
remaining notation is as in Fig.~\ref{fig:allN2LO}.
The red squares on the pion propagators  
indicate that for each diagram, the pion propagator
cancels  parts of the four-pion vertex expression,
as explained in the text.} 
\end{figure}

%%%%%%%%%%%%%%%%%%%%%%%%%%%%%%%%%%%%%%%%%%%%%%%%%%%%%%%%%%%%%%%%%%%%%%%%%
In order to illustrate the power counting of the loop diagrams with 
$\Delta$ in MCS we, 
as an example, discuss in detail the power counting for diagram type $\Delta$IV  in
Fig.~\ref{fig:DIV}.   
%%%%%%%%%%%%%%%%%%%% %%%%%%%%%%%%%%%%%%%%%%%%%%%%%%%%%%%%%%
First,  note that the  four-pion  vertex of the leftmost diagram 
in Fig.~\ref{fig:DIV}  can  be  rewritten  as  a  linear combination of the   three pion 
propagators  adjacent  to this  vertex  plus  a residual  vertex term\footnote{ 
While  the  first  three terms  depend explicitly on 
the  parameterisation (or ``gauge'') of  the pion  field,  
the  residual  term  is  
pion-gauge  independent \cite{subloops}.}  
%%%%%%% 
(see, e.g., appendix A.4 in  Ref.~\cite{NNLOswave}). 
This results in a separation of diagram  $\Delta$IV  in   four  parts:  for  the  diagrams   
$\Delta$IV  a-c the pion propagator cancels   corresponding parts of the  
four-pion vertex,  as  indicated by the red  square 
in   Fig.~\ref{fig:DIV},   while  diagram $\Delta$IV d  appears 
as  the residual part in this  separation. 
%
%
%%%%%%%%%%%%%%%%%%%%%%%%%%%%%%%%%%%%%%%%%%%%%%%%%%%%%%%%%%%%%%%%%%
%%%%%%%%%%%%%%%%%%%%%%%%%%%%%%%%%%%%%%%%%%%%%%%%%%%%%%%%%%%%%%%%% 
To estimate  the magnitudes of the amplitudes of these diagrams,  
we first remind the reader that  for the reaction $NN\to NN\pi$ close to threshold   
the initial nucleons have four-momentum  
$p_1^\mu = (m_\pi /2, \vec p \,)$ and  $p_2^\mu = (m_\pi /2, -\vec p \,)$  
with  $p=|\vec p \,|\approx \sqrt{\mpi m_N}$  (see  Fig. \ref{fig:DIV} for  the  notation). 
Secondly,  we  note that  the 
loop diagrams with  the  explicit $\Delta$ all involve  
the  $\Delta$-$N$ mass  difference  $\delta\sim p$  in the $\Delta$ propagator. 
The power counting for  the loop  diagrams also requires 
the inclusion of the integral measure
$l^4/(4\pi)^2$ where all components of the loop four-momentum $l$ 
are of order $\delta\sim p$, i.e. $v \cdot l \sim |\vec{l}| \sim p$.   
In addition to this integral measure, in the diagrams $\Delta$IVa--c  
one has to account for  one  $\Delta$ propagator ($\sim
1/(v\cdot l)\sim 1/p$),   three pion propagators ($\sim 1/(p^2)^3$),    
the $4\pi$ vertex ($\sim p^2/f_\pi^2$) and   
two $\pi N\Delta$ and one  $\pi NN$ vertices ($\sim (p/f_\pi)^3$).
Combining 
all these factors and using $4\pi f_\pi\sim m_N$, 
one obtains the order estimate for this diagram as follows 
\begin{eqnarray}
\frac{p^4}{(4\pi)^2} \, \frac{1}{p} \, \frac{1}{(p^2)^3} \, \frac{p^2}{f_\pi^2}\,  
\left(\frac{p}{f_\pi}\right)^3
\sim \frac{1}{\fpi^3}\frac{p^2}{m_N^2} \simeq \frac{1}{\fpi^3}\chi_{\rm MCS}^2.
\end{eqnarray}
This order estimate of diagrams $\Delta$IVa--c based on dimensional arguments 
should be compared with the corresponding estimate 
of a leading-order diagram for the $NN\to NN\pi$ 
reaction, namely the rescattering  diagram  with the Weinberg--Tomozawa vertex, 
 as  given  by  Eq.\eqref{treePC}.  
Thus, we  find that the diagrams  $\Delta$IVa--c   
in Fig.~\ref{fig:DIV}    start  to  contribute at  NLO.   
Meanwhile,  the  pion-gauge independent diagram $\Delta$IV~d  starts  to  contribute  at  
\NNLO{} only. 
%%%%%%%%%%%%%%%%% %%%%%%%%%%%%%%%%%%%%%%%%%%%%%%%%%%%%%%%
%
The reason is   that the  residual  pion-gauge independent  $4\pi$ vertex  is suppressed 
compared  to the leading $4\pi$ contributions. 
Finally,    notice  that   the   expression for diagram $\Delta$IV  contains more terms
 than the corresponding pure pion-nucleon  diagram IV.   
In the  pion-nucleon  case,  the contributions  similar to  type 
 $\Delta$IVb  and  $\Delta$IVc  are  strongly  suppressed  since  they  involve  only  the momentum scale of  the order of $\mpi$,
 as  explained in
Appendix A.4 in Ref.~\cite{NNLOswave}.

%%%%%%%%%%%%%%%%%%%%%%%%%%%%%%%%%%%%%%%%%%%%%%%%%%%%%%%%%%%%%%%%%%
\begin{figure}[t]
\includegraphics[scale=0.9]{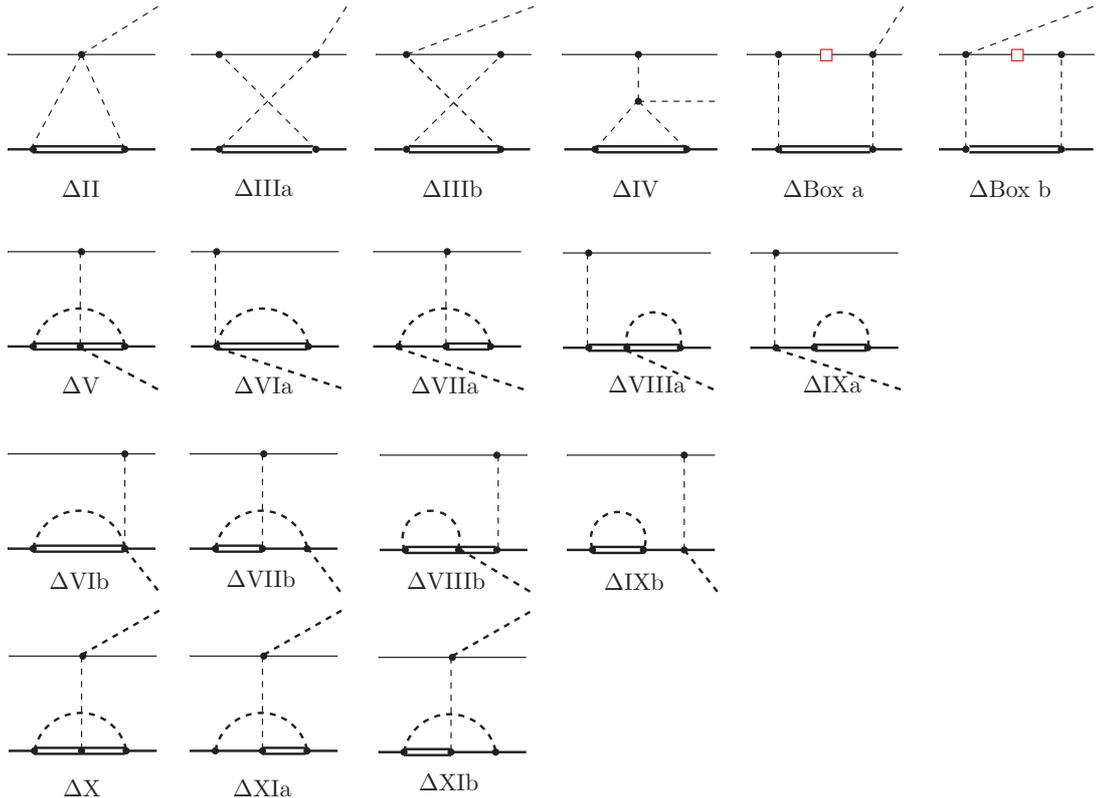}
\caption{\label{fig:allDLoops}
Loop diagrams with  the $\Delta$ degree  of  freedom contributing to 
s-wave pion production up to \NNLO{}.
Double lines denote the $\Delta$ propagator,  
remaining notation is as in Fig.~\ref{fig:allN2LO}. Again,
the red square on the nucleon propagator in the two box diagrams indicates 
 that the corresponding nucleon propagator cancels  parts of 
the Weinberg--Tomozawa $\pi N$ rescattering vertex leading to 
an irreducible contribution of the box diagrams 
as discussed in Sec.~\ref{sec:diagrams}. 
} 
\end{figure}
%%%%%%%%%%%%%%%%%%%%%%%%%%%%%%%%%%%%%%%%%%%%%%%%%%%%%%%%%%%%%%%%% 

\begin{figure}[t]
\includegraphics[scale=0.89]{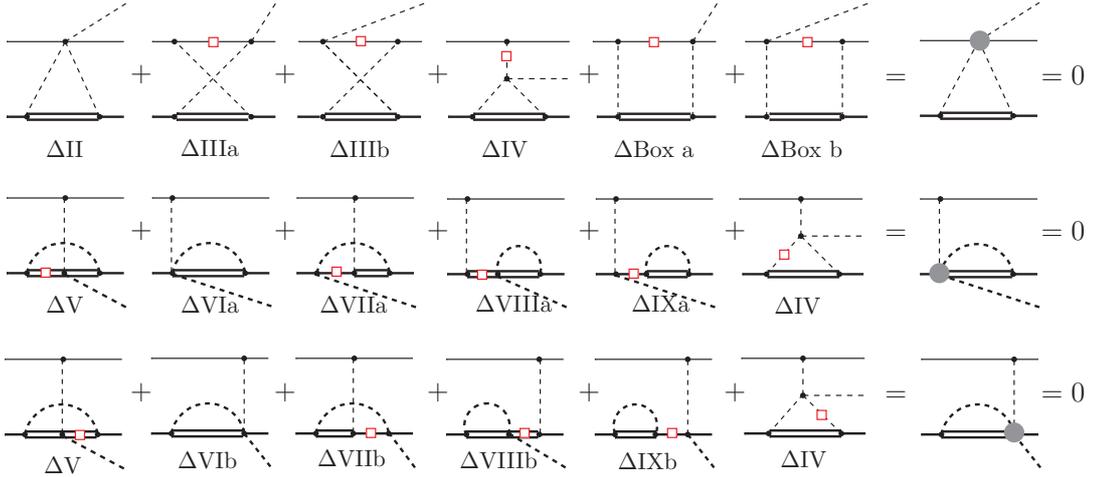} 
\caption{\label{delcanc}
Illustration of the cancellation pattern  among  the $\Delta$-loop contributions for  
different topologies shown on the   l.h.s. of each row. 
The red squares on the nucleon or $\Delta$ propagators indicate that for each diagram 
this nucleon or $\Delta$ propagator 
cancels  parts of the adjacent $\pi N$ or $\pi \Delta$ rescattering vertex. 
The red squares on the pion propagators  
indicate that for each diagram the pion propagator
cancels  parts of the four-pion vertex expression. 
These propagator cancellations generate 
the rightmost effective diagrams in each row where the  
effective vertices (blobs)  receive contributions from all the diagrams 
on the l.h.s. in the
corresponding row. 
The zero on  the very  r.h.s. in each row means that  
the sum of all diagrams on the l.h.s. 
contributes nothing to the s-wave pion  production  at  least  up to  \NNLO{}. 
}
\end{figure}

The loop  diagrams involving  $\Delta$  which  contribute  
to  s-wave pion  production up
to  \NNLO{}   are shown in  Fig. \ref{fig:allDLoops}.
In the  first row of  Fig.~\ref{fig:allDLoops}, we have
the two-pion exchange diagrams with topologies 
completely  analogous to the pion-nucleon $g_A^3$-graphs  in 
the second row in 
Fig.~\ref{fig:allN2LO}. 
The two-pion exchange diagrams in the first row 
of Fig.~\ref{fig:allDLoops} individually  start  to contribute  at NLO. 
However,  these NLO  diagrams cancel completely  in the sum  for 
the same reason as do the NLO pion-nucleon  ones in Fig.~\ref{fig:allN2LO}.  
In fact, it is relatively straightforward to show that,  
on the operator level, 
this cancellation of the NLO level diagrams is  independent of whether we
have a nucleon or a $\Delta$ propagator on the lower baryon line in
Figs.~\ref{fig:allN2LO} and \ref{fig:allDLoops}. 
Consequently, in MCS there are no 
contributions from these two-pion exchange diagrams at NLO. 
Moreover,  the \NNLO{}  contributions  of the diagrams in the first row 
in Fig.~\ref{fig:allDLoops} also  show  cancellation patterns among 
the diagrams absolutely  analogous to the purely pion-nucleon  case. 
In the  first  row  of
Fig.\ref{delcanc}, we demonstrate  graphically this systematic 
cancellation pattern of these diagrams   at  \NLO{}  and \NNLO{}, 
where a nucleon  (pion) 
propagator cancels parts of a Weinberg-Tomozawa $\pi N$ 
rescattering  (four-pion) vertex  
expression, rendering an effective diagram of topology like diagram $\Delta$II 
but with an effective three-pion-nucleon vertex which vanishes.  
It  should be mentioned  that   the  diagrams 
 in  the  first  row  of     Fig.~\ref{fig:allDLoops}  
also  obtain corrections  from 
higher-order  vertices  $\propto  1/m_N$ and $c_i$ (Delta-subtracted) from   
${\cal L}^{(2)}_{\pi\!N}$.  
Those  corrections,  however,  again  cancel  completely at \NNLO{} in a full 
analogy  to the cancellations among the  corresponding pion-nucleon loop  contributions,  
see  Ref.~\cite{NNLOswave}   and  the discussion  in  Sec.\ref{sec:loops}. 
The net sum of the \NNLO{} diagrams in the first row of 
Fig.~\ref{fig:allDLoops} receives
contributions only from   diagrams  
 $\Delta$IIIa and   $\Delta$IIIb,     
where  the  Weinberg-Tomozawa $\pi N$ vertices are on-shell, 
and a remnant of diagram  $\Delta$IV, the pion-gauge independent 
$\Delta$IVd shown in Fig.\ref{fig:DIV}. 
In other words,  the contributions  of  the diagrams $\Delta$IVa--c 
in  Fig.\ref{fig:DIV}  cancel against other diagrams, 
as indicated in Fig.~\ref{delcanc}, 
and only  $\Delta$IVd  with a residual part of the four-pion vertex remains.  

%%%%%%%%%%%%%%%%%%%%%%%%%%%%%%%%%%%%%%%%%%%%%%%%%%%%%%%%%%%%%%%%%%%%%%%%%%%%%%%
%%%%%%%%%%%%%%%%%%%%%%%%%%%%%%%%%%%%%%%%%%%%%%%%%%%%%%%%%%%%%%%%%%%%%%%%%%%%%%% 

In addition,  there are several new loop diagrams containing $\Delta$ propagators 
where one effectively has  a pion being exchanged between the two
nucleons,  see diagrams   $\Delta$V--$\Delta$XI in  Fig.~\ref{fig:allDLoops}.  
Surprisingly,  parts  of   diagrams $\Delta$V--$\Delta$IX  in
rows two and three  also undergo  significant cancellations.
Again,  as illustrated in rows two and three in Fig.~\ref{delcanc}, 
after  the  cancellations of   the  vertex  structures  with
the propagators,   some  parts  of   the   diagrams  $\Delta$V--$\Delta$IX  
and    $\Delta$IV acquire  the  effective  topology  of
the  diagrams  $\Delta$VIa and $\Delta$VIb.   
The  net  effect  of such  contributions again  
vanishes  at NLO and \NNLO{}  for  s-wave pion production 
as indicated on the r.h.s. in rows two and three of Fig.~\ref{delcanc}.  
After    the  cancellations,   only  those  parts  of  diagrams  
$\Delta$V--$\Delta$IX  remain,   
that   are  proportional  to the  on-shell part,   $ 2\mpi$, 
of  the $\pi \Delta-\pi\Delta$  
and  $\pi N-\pi N$  vertices.
Although the  dimensional  analysis estimate  indicates  that these  
residual  contributions  
are naively of \NNLO{} in MCS,  
most of the  \NNLO{} amplitude expression   vanishes due to  loop angular integration.  
For  example,  for   diagrams $\Delta$VII    the  numerator    of the integrand 
$\propto  2\mpi\  \Sd\cdot (p_1- p_1')  (\Sd\cdot l)$   is   
odd  with  the loop  momentum $l$  yielding  the  vanishing  contribution at  \NNLO{}.  
As a consequence of these cancellations   
almost all loop diagrams  in    rows two and three  in Fig.~\ref{fig:allDLoops}   
do not contribute to the s-wave \NNLO{} pion-production amplitude. 
Only diagram $\Delta$V in the second row yields a   non-vanishing  \NNLO{}
contribution  from  the  loop diagrams  in rows two and three, 
where again  only the  on-shell part of the 
$\pi \Delta-\pi\Delta$ rescattering vertex  in diagram $\Delta$V remains.

Finally, the three one-pion-exchange $\Delta$ loop diagrams 
in the  last  row  of   Fig.~\ref{fig:allDLoops}, 
which  have to be taken into account at  \NNLO{}, 
contribute only  to the  renormalization of  $g_A$  at \NNLO{}, see the next subsection. 
Furthermore, the  time-dependent
Weinberg--Tomozawa $\pi N$ vertex in these three diagrams appears  on-shell  
as discussed in the text after Eq.(\ref{eq:pipivert}).

%%%%%%%%%%%%%%%%%%%%%%%%%%%%%%%%%%%%%%%%%%%%%%%%%%%%%%%%%%%%%%%%%%%%%%%%%%%%%%%%%%%%%% 
\subsection{Regularization of UV divergences and renormalization}
\label{sec:renorm2}
%%%%%%%%%%%%%%%%%%%%%%%%%%%%%%%%%%%%%%%%%%%%%%%%%%%%%%%%%%%%%%%%%%%%%%%%%%%%%%%%%%%%%%

%%%%%%%%%%%%%%%%%%
\begin{figure}[t]
\includegraphics[scale=0.47]{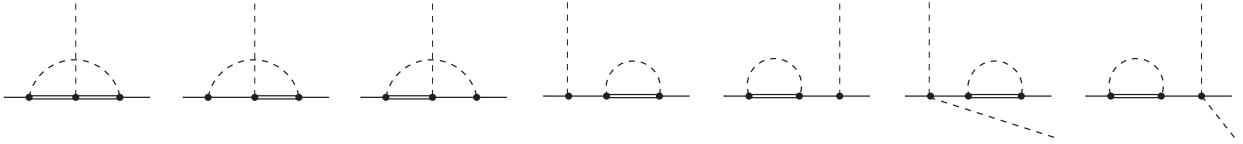} 
\caption{\label{renormD}
Various renormalization diagrams  relevant at N$^2$LO.
The first three diagrams contribute to the renormalization correction of  
the $\pi N$  coupling constant $g_A$, whereas the last four diagrams renormalize 
the  nucleon wave  function using the  leading  vertices in  the theory with explicit  $\Delta$  degrees  of freedom.}
\end{figure}
%%%%%%%%%%%%%%%%%%

The loop diagrams  with explicit $\Delta$  are  
UV-divergent at \NNLO{}. 
These $\Delta$ loop diagrams require that   
the couplings and masses appearing in the Lagrangian  should be renormalized.  
In particular,  up  to \NNLO{}  in  MCS  we concentrate on two  relevant
renormalization  corrections, 
namely the ones to the  $\pi N $  coupling constant  $g_A$  and 
the   nucleon  wave  function renormalization 
factor $Z_N$,  as   shown  in Fig.\ref{renormD}. 
These renormalization corrections  of order  
$\delta^2/\Lambda_\chi^2 \sim \chi_{\rm MCS}^2$
were   evaluated in Ref.~\cite{Fettes,Fettes2}  
  for $\pi N$ scattering with explicit
$\Delta$ in the loop using dimensional  regularization. 
We  confirm  the results  of Ref.~\cite{Bernard:1998gv,Fettes,Fettes2}:  
\begin{eqnarray}
    Z_N &=& 1 + 2 (d-2) \gpind^2 \frac{\delta }{\fpi^2}\jpid
        + {\cal O}\left( \frac{\mpi^2}{\Lambda^2} \right), 
 \\ 
	\mathring{g}_A &=&        \nonumber
 \ga 
	    + \frac{10 (6+d-4d^2 +d^3)}{9 (d-1)^2} \, \gone \, \gpind^2  \,  
        \frac{\delta }{\fpi^2} \, \jpid 
	    + \frac{16 (d-2)}{3 (d-1)^2}\, \ga \, \gpind^2  \, \frac{\delta }{\fpi^2}\, \jpid
	    + {\cal O}\left( \frac{\mpi^2}{\Lambda^2} \right) \; ,
        \label{garenorm} 
\end{eqnarray} 
where $\mathring{g}_A$
%%%%%%% 
is the bare axial coupling,  $\gpind=h_A/2$ and      
$g_1= 9/5 g_A$ is the $\pi \Delta\Delta$  coupling  constant. 
%%%%%%%%%%%%%%%%%%%%%%%
In order to  calculate the  production operator  up to  NLO  it  suffices to   use  
$Z_N =1 $  and  $\mathring{g}_A =\ga $.  
However,  in a theory  with  explicit  $\Delta$  degrees   of   
freedom, renormalization  corrections  to the tree-level  diagrams at LO 
in Fig.~\ref{fig:allN2LO}  generate  \NNLO{}  contributions. 
At  \NNLO{}, the nucleon  fields ($N$) in  
the Lagrangian 
must  be  renormalized, i.e.,   $N \to  N \sqrt{Z}$,  
and, similarly, for the axial constant $\mathring{g}_A \to g_A$   
(i.e., $\mathring{g}_A$  deviates  from the  physical  value),   
due  to the loop corrections with  explicit  $\Delta$ in Eq.~\eqref{garenorm}.  
%%%%%%%
The explicit evaluations  of the diagrams in Fig.~\ref{fig:allDLoops} 
reveal that the  contributions  of  diagrams  
$\Delta$X and  $\Delta$XI   
reproduce the \NNLO{}  correction to the  tree-level rescattering  
diagram in Fig.~\ref{fig:allN2LO}  
due to renormalization of  $\mathring{g}_A$  and $Z_N$ 
(see first  five  diagrams  in Fig.\ref{renormD}),  
meaning  that  at \NNLO{} there is no genuine contributions  
of diagrams  $\Delta$X and  $\Delta$XI  in MCS.

The  \NNLO{} contribution from  $Z_N$ to the  WT  vertex 
(the last two  diagrams  in Fig.\ref{renormD}), 
is included  in  the
rescattering  operator  together  with  the  residual  contributions
of the  diagrams  $\Delta$III, $\Delta$IV and  $\Delta$V and  
gives non-vanishing  correction     at   \NNLO{}.   
The individual non-vanishing contributions  of the    
$\Delta$ loop diagrams  in Fig.~\ref{fig:allDLoops},  
 expressed  in terms of four well-known  scalar integrals,   
$\jpid$, $\ipipi$, $\jpipid$, and $\jpipind$, which are 
defined in  appendix \ref{sec:appbasicint},  
Eqs.~(\ref{JpiD})--(\ref{JpipiND}), read 
\begin{eqnarray}
 i M_{\Delta \rm III(a+b)} &=& i (SI_1) \frac{1}{(d-1)(d-2)} 
        \left\{ \intI  - \frac12 \intJ  + \intT + \frac14 \intL \right\} 
        \nonumber
\\             &+& i (SI_2)\, \frac{d-2}{d-1} 
        \left\{   \frac12 \intI - \frac12 \intJ + \frac12 \intT + \frac14 \intL \right\}\, ,
        \nonumber
\\
i M_{\Delta \rm IV}   &=& i (SI_2)\, \frac{d-2}{d-1} 
        \left\{ 
            \left( 2 - \frac{1}{d-1} +  4\, \frac{\delta^2}{k_1^2}\right) \intI
            -4 \frac{\delta^2}{k_1^2} \intJ 
            + \left( 2 +  4 \frac{\delta^2}{k_1^2}\right) \intT
        \right\} \, ,
        \nonumber
\\ 
i M_{\Delta \rm V} &=& i (SI_2)\, (d-2) \left\{ 5
        \frac{\delta^2}{k_1^2} \intJ \right\} \, ,
        \nonumber
\\
i M_{\Delta}^{Z_N} &=& i (SI_2)\, (d-2) \left\{ -3 \frac{\delta^2}{k_1^2} \intJ \right\} \, ,
        \label{eq:MDelta}
\end{eqnarray}
where the two spin-isospin operators in Eq.(\ref{eq:MDelta})   are:  
\begin{eqnarray}
	(SI_1) &=& (-i) \frac{\gpind^2}{\fpi^5}  \ga 
        \left\{ \left( \frac23 \tau_1^a - \frac13 \tau_2^a \right)
	    4 \left[ S_{2\mu}, \, S_{2\nu}\right] S_1^\nu k_1^\mu \vdotq \quad+  (1
        \leftrightarrow 2) \right\}, 
        \nonumber
\\
	(SI_2) &=& (-i) \frac{\gpind^2}{\fpi^5}  \ga\,  \frac{i}{3}  
        \left\{ (  \bm\tau_1 \times \bm \tau_2)^a 
	    S_1 \cdot  k_1 \vdotq  \quad+ (1
        \leftrightarrow 2) \right\}. 
\end{eqnarray} 
The four different loop integrals in Eq.~(\ref{eq:MDelta})  
can be characterized in the following manner. 
The  integral  $\jpipid$, Eq.~(\ref{JpipiD}),  contains  two pion propagators and
a  $\Delta$ propagator whereas  the integral  $\jpipind$,  Eq.~(\ref{JpipiND}),  
has  an  additional  nucleon propagator.  
Furthermore, we note that both  of these integrals  are  UV   finite. 
Meanwhile,  the integrals  $\jpid$ and $\ipipi$  contain UV divergences and,  
similar  to  the  pion-nucleon loops,  some  of these  divergencies  
can be absorbed  by  the
five-point $NN\to NN\pi$ contact  term,  see  the  last  row in Fig.\ref{fig:allN2LO}. 

Before we present the final transition amplitude contribution for s-wave pion production 
from $\Delta$ loop diagrams, there is one issue which deserves attention. 
In a  theory  containing a ``heavy"  resonance $\Delta$,  
it  is not  sufficient to  require  just  the  cancellation of the UV  divergent  
terms with the corresponding LECs. 
The integrals  $\jpid$ and $\ipipi$ in  Eq.~\eqref{eq:MDelta}
which are multiplied by the factor  $ \delta^2/k_1^2$ 
pose an additional problem.  
Such polynomial behavior  would give       divergences                           
if the $\Delta$-resonance was infinitely heavy, i.e., if  $\delta \to \infty$.
Therefore,    to find the most natural finite values of the renormalized LECs,   
the explicit ``decoupling''
renormalization scheme  was  introduced~\cite{Appelquist}.
In such a scheme, the finite parts of  LECs are 
chosen  such that the renormalized contribution 
from diagrams with $\Delta$ loops vanish in the limit $\delta \to \infty$.
It  is   demonstrated in  Appendix \ref{intdecoupl} that  the following combinations 
of loop integrals (up to \NNLO{} in MCS) do vanish when $\delta \to \infty$: 
\begin{eqnarray}
\label{intcomb1}
		&&\intL, 
		\\
\label{intcomb2}
		&&\left( \intI  + \frac12 \intJ + \intT + \intC{2} \right),
		\\
\label{intcomb3}
		&&\frac{\delta^2}{k_1^2} \left( \intI  + \frac12 \intJ + \intT + \intC{2}\right) 
	    	- \frac{1}{12} \left(  \intI + \frac12 \intJ + \frac13 \intC{2} \right).
\end{eqnarray}
We find that the combination of the two integrals 
$\jpid$ and $\ipipi$ in Eqs. (\ref{intcomb1})-(\ref{intcomb3}),  
$\ipipi + \frac{1}{2\delta}\jpid $,  
cancels the  UV divergences of  the  individual integrals, 
as  proven  at the  end of the Appendix \ref{intdecoupl}. 
Hence,  Eqs.~\eqref{intcomb1}-\eqref{intcomb3} are all  UV  finite  
and  vanish  when $\delta \to \infty$.

Rewriting the sum of the amplitudes  $M_{\Delta \rm III(a+b)},
M_{\Delta\rm  IV},  M_{\Delta \rm V}$ and $ M_{\Delta}^{Z_N}$  from
Eq.~(\ref{eq:MDelta}) in terms of the integral  combinations
\eqref{intcomb1}--\eqref{intcomb3},  
we obtain the following transition amplitude from the $\Delta$ loop diagrams 
\begin{eqnarray}
\nonumber 
    &&  \hspace*{-0.5cm} i M_{\Delta\text{-loops}}^{\text{\NNLO{}}} 
          = \frac{\ga \gpind^2}{\fpi^5} \vdotq \,
	    \tau_{+}^a \left( i \varepsilon^{\alpha \mu \nu \beta} v_\alpha k_{1\mu} 
               S_{1 \nu} S_{2 \beta} \right) 
	    \\
             \nonumber
   \times&&  \hspace*{-0.4cm}
	    \Bigg\{ 
	    	    \frac29 \left( \intI + \frac12 \intJ + \intT + \intC{2} \right) 
	    	  + \frac{1}{18} \intL 
		- \left[ \frac{2}{3(d-1))} \intJ \right] + \tilde{\cal A}^{\Delta}_\text{CT}
	    \Bigg\} 
	    \\ 
             \nonumber
   +&& \hspace*{-0.3cm} \frac{\ga \gpind^2 }{\fpi^5} \vdotq \, \tau_{\times} (S_1 + S_2) \cdot k_1 
   \\\nonumber
        \times&&\hspace*{-0.4cm}
        \Bigg\{
 	    	\frac{5}{9} \left(  \intI + \frac12 \intJ + \intT + \intC{2} \right) 
	    	+ \frac{1}{18} \intL   - \frac{2}{27} \left(  \intI + \frac12 \intJ + \frac13 \intC{2} \right) 
	    	\\
	    	+&&\hspace*{-0.45cm} \frac{8}{9} \frac{\delta^2}{k_1^2} \left( \intI + \frac12 \intJ  + \intT 
            + \intC{2}\right) 	    	- \left[ \frac{(d-2)}{3(d-1)} \left(\frac{19}{12} \intJ +  \intC{5}\right)  \!\right]  
          +  \tilde{\cal B}^{\Delta}_\text{CT}
       \! \Bigg\}. 
\label{eq:MDfinal}
\end{eqnarray} 
Additional terms,  that do not vanish at the large $\delta$ limit, 
 shown  in  the  square  brackets  in Eq.~\eqref{eq:MDfinal},
are short-ranged and  are cancelled in the    
amplitude  expression  by  the parts  of    
the five-point contact terms  $\tilde{\cal A}_\text{CT}^{\Delta}$ and 
$\tilde{\cal B}_\text{CT}^{\Delta}$ due  to the  explicit  $\Delta$.   
In other  words,  the  decoupling  condition   fixes  the  magnitude  of 
the five-point  contact interactions due to the explicit $\Delta$,  
$\tilde{\cal A}^{\Delta}_\text{CT}$ and $\tilde{\cal B}^{\Delta}_\text{CT}$,  
up  to  higher  order  terms.  
In fact, by defining the five point contact terms as
\begin{eqnarray}\label{fivepointpl}
	\tilde{\cal A}^{\Delta}_\text{CT}  &=&  \frac{2}{3(d-1)} \intJ +  
       O\left(\frac{\mpi^2}{\delta^2}\right),
	\\\label{fivepointcr}
	\tilde{\cal B}^{\Delta}_\text{CT}  &=& 
	 \frac{(d-2)}{3(d-1)} \left(\frac{19}{12} \intJ + 
       \intC{5} \right)+ O\left(\frac{\mpi^2}{\delta^2}\right)  \, ,
\end{eqnarray}
we obtain the fully renormalized, finite  $\Delta$ loops amplitude contribution, which
satisfies  the  decoupling condition, to  s-wave pion  production at  \NNLO{}:
%%%%%%%% 
\begin{eqnarray}
\label{delsymmampstart}
\nonumber
		&& \hspace*{-0.5cm} i M_{\Delta\text{-loops}}^{\text{\NNLO{}}} 
       = \frac{\ga \gpind^2}{\fpi^5} \vdotq \,
	    \tau_{+}^a \left( i \varepsilon^{\alpha \mu \nu \beta} 
                v_\alpha k_{1\mu} S_{1 \nu} S_{2 \beta} \right) 
	    \\
          \nonumber
	 \times&&  \hspace*{-0.4cm} 
	    \Bigg\{ 
	    	    \frac29 \left( \intI + \frac12 \intJ + \intT + \intC{2} \right) 
	    	  + \frac{1}{18} \intL   
	    \Bigg\} 
	    \\
           \nonumber
   +&& \hspace*{-0.3cm}      \frac{\ga \gpind^2 }{\fpi^5} \vdotq \, \tau_{\times} 
          (S_1 + S_2) \cdot k_1 
   \\
      \nonumber\label{delsymmampend}
      \times&&\hspace*{-0.4cm} 
        \Bigg\{
 	    	\frac{5}{9} \left(  \intI + \frac12 \intJ + \intT + \intC{2} \right) 
	    	+ \frac{1}{18} \intL   
	    	\\ 
	+&&\hspace*{-0. cm}      \frac{8}{9} \frac{\delta^2}{k_1^2} \left( \intI  + \frac12 \intJ 
                            + \intT + \intC{2}\right) 	- \frac{2}{27} \left(  \intI + 
          \frac12 \intJ + \frac13 \intC{2} \right) 
	        \Bigg\}. 
\end{eqnarray} 
This expression should be added to the finite s-wave production operators 
presented in Sec.~\ref{sec:nucl}.

%%%%%%%%%%%%%%%%%%%%%%%%%%%%%%%%%%%%%%%%%%%%%%%%%%%%%%%%%%%%%%%%%%%%%%%%%%%%%%%%%%%%%%
\section{Comparison of  the   pion-nucleon   and  $\Delta$  loop   contributions}
\label{sec:Delta_N}
\begin{figure}[t]
\includegraphics[scale=1.1]{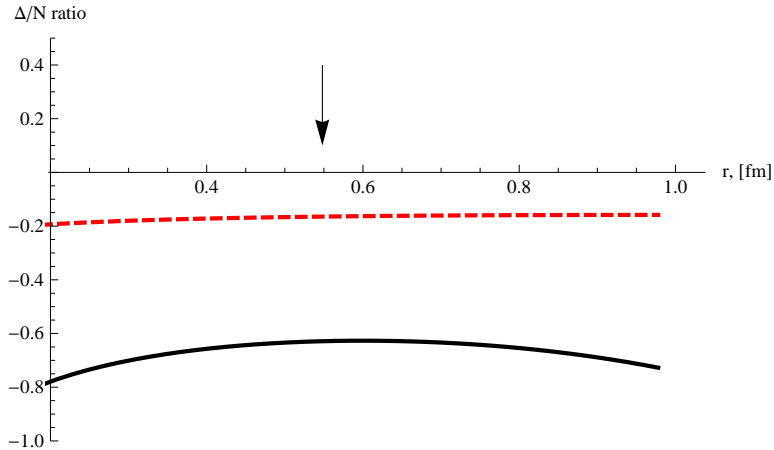} 
\caption{\label{ratioDN}
The ratio  of  the $\Delta$  contributions  to the pion-nucleon ones  for
$\bm\tau_+$ (red dashed curve)   and  $\bm\tau_{\times}$ (black solid curve)
channels. The arrow  indicates  $r\sim 1/p=1/\sqrt{m_\pi m_N}$.}
\end{figure}

%%%%%%%%%%%%%%%%%%%%%%%%%%%%%%%%%%%%%%%%%%%%%%%%%%%%%%%%%%%%%%%%%%%%%%%%%
In Ref.~\cite{NNLOswave} (see also discussion in Sec.~\ref{sec:nucl})  
we argued  that the  long-range scheme-independent  part
of   the  pion-nucleon  loops  at  \NNLO{} is sizable  and   could  
resolve  the problem  with the description of  pion  production data
in the  neutral  channel, $pp \to pp \pi^0$.   
We now add the long-ranged $\Delta$ contribution. 
First,  we note  that 
the spin-isospin structure of the    $\Delta$-loops  in Eq.~(\ref{delsymmampstart}) 
is exactly
the  same as for the  pion-nucleon  case  in  Eq.~\eqref{eq:Mga3final}.  
Meanwhile,  the
dimensionless  integrals are different  and   the 
coefficients  in front  of the spin-isospin operators  also differ.  
We want to compare the resultant amplitudes  from the 
nucleon and $\Delta$ loop diagrams for $NN$   
relative distances  relevant  for
pion production,  i.e.  for $r\sim 1/p \simeq 1/\sqrt{m_\pi m_N}$. 
In order to separate the long-range scheme-independent  contributions 
of  the  $\Delta$-loop expressions from  the short-range ones 
in a transparent manner, we make  
a  Fourier  transformation of our expressions.  
The Fourier  transformation of 
a short-range (constant) contribution   gives   a  
$\delta$-function,   $\delta({\bf r})$,   which  does
not  influence the long-range  physics  of interest and 
we therefore ignore this contribution in this section.    
The  Fourier  transformation  of the long-range part of the 
loop integrals is evaluated numerically  
as follows:
\begin{eqnarray}
	I(r)= \int  \frac{d^3 k}{(2\pi)^3} e^{i {\bf k r}} \, I(k)\, e^{-k^2/\Lambda^2} \, . 
\end{eqnarray}
Here,  the  regulator $e^{-k^2/\Lambda^2}$    is  used  
in order to minimize the influence of the 
large momenta  in  the  loop integrals, denoted by $I(k)$ for short.  
We have verified  that  for
$\Lambda>$2 GeV this regulator   does  not
affect the  results  in  the long-range   region  of  $r\sim 1/p$.
Specifically,  we  Fourier  transform   the   integral
combinations  in the  curly   brackets  in  Eq.~\eqref{delsymmampend}
(multiplied by $\gA \gpind^2$)
corresponding to   $\tau_+$ (neutral) and  $\tau_{\times}$ (charged)  channels. 
We  compare the resulting Fourier transformed amplitudes  with  
the Fourier  transformed amplitudes of  the  corresponding 
pion-nucleon  contributions in Eqs.~\eqref{eq:Mga1final} and \eqref{eq:Mga3final},  
$-2 \gA^3 I_{\pi\pi}$ and  $(-19/24 g_A^3+1/6 g_A)I_{\pi\pi} $, respectively. 
The  ratio  of the  $r$-space $\Delta$ loop 
contributions  to those  of the   nucleon   is  shown  in
Fig.~\ref{ratioDN}.  
One can see  that in the  neutral  channel,  the
long-range part of the $\Delta$ contribution   constitutes  
less  than  20\% compared 
to  the  pion-nucleon loop amplitude.  
This  can be 
understood by the
specific  combination of the  coefficients 
for the  spin-isospin operator 
in the case of the  $\Delta$-resonance amplitude, 
which results in a suppression  by  almost one order of magnitude. 
Therefore,  the  conclusion  of  Ref.~\cite{NNLOswave},   
regarding the  importance
 of  the pion-nucleon  loops in explaining the  neutral  pion  production, 
appears to be  only slightly modified by the  $\Delta$ loop contributions. 
Regarding the  charged channel  the  $\Delta$-loop contribution  to the 
s-wave pion-production amplitude is  almost of
the same  magnitude (roughly 60\%) but of opposite sign compared to the pion-nucleon one. 
The net loop amplitude from the nucleon and $\Delta$ loop diagrams is therefore 
not as important as in the neutral channel.   

The pattern that emerged from the loops is therefore exactly what is necessary 
to quantitatively describe the data on both $pp\to pp\pi^0$ and $pp\to d\pi^+$
very near threshold: while in the former reaction there persists a huge discrepancy
between data and the chiral perturbation theory calculation to NLO, in the latter
 at NLO the description is already quite good~\cite{towards}. 
In line with this we now find that due to large cancellations amongst the pion-nucleon
and the Delta loops the N$^2$LO corrections from the loops are small in the charged
pion channel. On the other hand, this cancellation is by far not that efficient in the 
neutral pion channel leading to a significant loop contribution. In combination with
the observation that in the neutral pion channel the leading order diagrams are
suppressed both kinematically as well as dynamically~\cite{hanhart04}, this provides
a dynamical reason of why it was so much harder to understand phenomenologically
the neutral pion production compared to the charged pion production.

%%%%%%%%%%%%%%%%%%%%%%%%%%%%%%%%%%%%%%%%%%%%%%%%%%%%%%%%%%%%%%%%%%%%%%%%
\section{SUMMARY AND CONCLUSIONS} 
\label{sec:outlook}
%%%%%%%%%%%%%%%%%%%%%%%%%%%%%%%%%%%%%%%%%%%%%%%%%%%%%%%%%%%%%%%%%%%%%%%%%

We have investigated the pion production operator in \NNNNpi{} near threshold 
within  chiral EFT. Specifically,  by explicit inclusion of the Delta-isobar 
and the $1/\mN^2$-corrections we have calculated the pion production 
operator to next-to-next-to-leading order (\NNLO{}) in the 
momentum counting scheme (MCS). The
MCS keeps track of the large initial momenta necessary in order to produce a
pion from two nucleons. 
The MCS approach accounts for the scales encountered in this reaction,
namely the pion mass, the intermediate momentum scale $p \sim \sqrt{\mpi \mN}$
compared to the hadronic scale $\Lambda_\chi$.
An understanding of such a two-scale problem is important also for other systems:
e.g. to formulate a power counting for the decays of heavy charmonia via meson loops 
it is necessary to keep track of various energy and momentum scales~\cite{charmloops}.

 While all loops cancel at \NLO{}, at \NNLO{} there is a finite remainder of the  $\Delta$ loop  contributions 
which is of the same order of magnitude as its purely pion-nucleon counter part
calculated in Ref.~\cite{NNLOswave}. 
This finding confirms the conjecture that the new scale introduced 
by the Delta-nucleon mass splitting should
be treated at the same order as the one from the initial momentum.

It is encouraging to observe that the sum of pion-nucleon and 
Delta loops shows exactly the pattern
required by the data. 
In the channel where there is an isoscalar 
$NN$ pair in the final state, there is a 
large cancellation between the \NNLO{} Delta loops and the pion-nucleon ones 
as indicated in Fig~\ref{ratioDN}. 
Note that in this channel the \NLO{}
calculation of Ref.~\cite{towards} was already close to reproducing the data. 
On the other hand, in the
channel where the final state $NN$ pair is in an isovector state 
the Delta loop contributions are much reduced compared to the pion-nucleon loops
as seen in Fig~\ref{ratioDN}. 
This means that the sum of Delta- and nucleon-loop contributions 
is sizable and it is precisely in this channel
where earlier calculations revealed huge deviations from data.

This indicates that the large quantitative difference 
between the two near threshold amplitudes 
${\cal A}$ and ${\cal B}$ in Eq~\eqref{eq:Mthr} can
be well understood on dynamical grounds  based on nucleon, 
Delta and pion degrees of freedom. 
Meanwhile,
the short-ranged physics contribution is 
parametrized by local contact terms in effective field theories, 
and  these LECs might
well give similar contributions in the two channels. 
However, in order to make this finding more
quantitative, a convolution with proper $NN$ wave functions is necessary. 
Formally, in a consistent chiral EFT calculation 
one should derive the transition operators and the initial 
and final nucleon wave functions from the same effective Lagrangian.
This has not yet been carried out and is beyond the scope of this work. 
For pragmatic reasons 
we  will consider the hybrid approach
where, as in this work, the transition amplitudes are evaluated in ChPT, 
whereas the nucleon wave functions 
are generated by modern phenomenological NN potentials. 
As a next step,  the
calculation of observables in the hybrid approach 
and the comparison with experimental data will 
be presented in a subsequent work.
The  intrinsic scheme-dependence  inherent to the hybrid approach also needs to  be  quantified.

Our work provides an important step forward towards understanding  the  important class
of reactions, namely the first inelastic $NN$ reactions --- $NN\to NN\pi$. 
Especially, it confirms the observation of Ref.~\cite{NNLOswave}, 
where it was pointed out 
that the long-range physics is not described  properly in earlier phenomenological 
calculations~\cite{Lee,HGM,Hpipl,jounicomment,eulogio,unsers,unserdelta}.   
Indeed,   none of the loop contributions  with  nucleons and Delta
that survived the significant
cancellations, like the pion crossed boxes (cf. diagrams IIIa and IIIb in   
Figs.~\ref{fig:allN2LO} and 
~\ref{fig:allDLoops}), 
were included in these works.
In addition, similar to cancellations among the pion-nucleon loop terms, also all the Delta loops  that can be associated
  with  scalar-isoscalar $\pi N$ interaction in the pion rescattering term cancel. 
This cancellation is in conflict with the claims of Refs.~\cite{eulogio,unsers}, 
where these
diagrams provided the essential contributions necessary to describe the data.

%%%%%%%%%%%%%%%%%%%%%%%%%%%%%%%%%%%%%%%%%%%%%%%%%%%%%%%%%%%%%%%%%%%%%%%%%%%%%%%%%
\section*{Acknowledgments}

This work was
supported in part by funds provided by 
the EU HadronPhysics3 project ``Study
of strongly interacting matter'', the European Research Council
(ERC-2010-StG 259218 NuclearEFT),  DFG-RFBR grant (436 RUS 113/991/0-1),
and the National Science Foundation Grant No. PHY-1068305. 

%%%%%%%%%%%%%%%%%%%%%%%%%%%%%%%%%%%%%%%%%%%%%%%%%%%%%%%%%%%%%%%%%%%%%%%%%%%%%%%%%%%%%%%%%%%%%
\appendix
%%%%%%%%%%%%%%%%%%%%%%%%%%%%%%%%%%%%%%%%%%%%%%%%%%%%%%%%%%%%%%%%%%%%%%%%%%%%%%%%%%%%%%%%%%
\section{Reaction amplitude}
\label{convol}
%%%%%%%%%%%%%%%%%%%%%%%%%%%%%%%%%%%%%%%%%%%%%%%%%%%%%%%%%%%%%%%%%%%%%%%%%%%%%%

In the most general case, an amplitude corresponding to the matrix element
of a particular production and/or absorption operator
between two-nucleon states with given initial ($j,l,s$) and final ($j',l',s'$) total
angular momentum of a 
nucleon pair,  its orbital momentum and total spin\footnote{In order to
unambiguously specify the partial wave, 
the pion angular momentum should, in general, also be given. We, however, omit it
since it is only the $s$-wave pion production 
that is  considered here.}
is written as
\begin{eqnarray}
 \!\! \!\! \mathcal{M}^\mathrm{full}[jls,j'l's']\!\!  =\!\! 
\maM^\mathrm{prod}[jls,j'l's']\!+\!\maM^\mathrm{FSI}[jls,j'l's']
\!+\!\maM^\mathrm{ISI}[jls,j'l's']
\!+\!\maM^\mathrm{ISI+FSI}[jls,j'l's'].  \
\label{suma}
\end{eqnarray}
Here $\maM^\mathrm{prod}[jls,j'l's'](p,p')$, with  $p$ and $p'$ being the 
initial and final nucleon relative momenta,  
 stands for the   $NN\to NN\pi$  production amplitude
where  there is no $NN$ interaction in the initial and in the final state, and 
$\mathcal{M}^\text{FSI}$, $\mathcal{M}^\text{ISI}$, $\mathcal{M}^\text{ISI+FSI}$
refer to the amplitudes with final state, initial state, and both final and
initial state interaction included, in order. 
In this equation we imply that the spin-angular part (as well as the isospin
part) of the amplitudes are factored out. 
Note that since there is a third particle that carries angular
momentum, the pion, the total angular momentum $j$ of the initial two-nucleon
state can be different 
from that of the final two-nucleon state, $j'$. 
Obviously, the total angular
momentum of the final 
particles equals that of the initial one.
The pion-production amplitude for off-shell kinematics ($\tilde\maM^\mathrm{prod}$) 
  includes,  in addition to $\maM^\mathrm{prod}$, the  single-nucleon (the direct diagrams)
production operator Eq.~\eqref{Mdir}  as well as  the tree-level operators 
involving $N \Delta$ intermediate states Eq.~\eqref{Mdel} 
(see  Fig.~\ref{fig:treeDelta} for the corresponding diagrams).
The other amplitudes in Eq.~\eqref{suma} are given by the following formulae:
\begin{eqnarray}\label{MFSI}
\maM^\mathrm{FSI}[jls,j'l's']\!\!&=& \!\!
C_{NB}\sum_{l''\!,s''}\int\frac{d^3q}{(2\pi)^3}\frac{\tilde\maM^\mathrm{prod}[jls,j'l''s''](p,q)\ 
\maM_{NB}[j',l''s'',ls](q,p')}%
{4m_{1'}m_{2'}\big[{q^2}/{(2\mu_{1'2'})}-(E^\prime-\delta)-i0\big]},\label{fsi}\\\label{MISI}
 \maM^\mathrm{ISI}[jls,j'l's']\!\!&=&\!\!C_{NB}
\sum_{l''\!,s''}\int\frac{d^3q}{(2\pi)^3}
\frac{\maM_{NB}[j,ls,l''s''](p,q)\ \tilde\maM^\mathrm{prod}[jl''s'',j'l's'](q,p')}%
{4m_1m_2\big[{q^2}/(2\mu_{12})-(E-\delta)-i0\big]},\label{isi}\\\label{MISIFSI}
\maM^\mathrm{ISI+FSI}[jls,j'l's']\!\!&=&\!\!C_{NB}
\sum_{l''\!,s''}\sum_{l'''\!,s'''}
\int\frac{d^3q}{(2\pi)^3}\frac{d^3\ell}{(2\pi)^3}\nonumber\\
& &\hspace{-0.16\textwidth}\times\frac{\maM_{NB}[j,ls,l''s''](p,q)\
  \tilde\maM^\mathrm{prod}[jl''s'',j'l'''s'''](q,\ell)\
  \maM_{NB}[j',l'''s''',l's'](\ell,p')}% 
{4m_1m_2\big[{q^2}/(2\mu_{12})-(E-\delta)-i0\big]\cdot4m_{1'}m_{2'}
  \big[\, {\ell^2}/({2\mu_{1'2'}})-(E^\prime-\delta)-i0\big]}\,,\label{fsiisi}%\\ 
\end{eqnarray}
where $m_{1,2}$ ($m_{1',2'}$) are the masses of the particles in the
intermediate state that are related via the $NN$ interaction  to 
the initial (final) state, $\mu_{12}$
($\mu_{1'2'}$) are the corresponding reduced masses, $E$ ($E^\prime$) is the
energy of the initial (final) two-nucleon state in its center-of-mass frame,
$\maM_{NB}[j,l_i s_i,l_f s_f]$ is the nucleon-baryon $NN-NB$ or  vice versa
 half-off shell $\maM$-matrix (with $B$ standing for $N$ or $\Delta$)
corresponding to a transition from the state $(jl_i s_i)$ to the state $(jl_f
s_f)$, and the sums are over all the intermediate states with given $j,\ j'$,
$l$, $l^\prime$, $s$, and $s^\prime$.  
We use the following relation between the
$\maM$-matrix and the commonly used $\mathcal T$-matrix:
$\maM_{NB}=-8\pi^2\sqrt{m_1m_2m_{1'}m_{2'}}~\mathcal T$, 
where the $m_i$ are the masses 
of interacting particles. 
Furthermore,  in the formulae  given above the coefficient   $C_{NB}$     
equals  to  1   for NN  intermediate  states  and   $2/\sqrt{2}$  in the  case  
when one  of  the  intermediate states (either  initial or final)  
contains  $\Delta$\footnote{Note that  keeping  $\Delta$  both 
in the initial  and  final states  constitutes  a higher order effect,  
as  discussed in Sec.\ref{sec:Delta_reduce}.}.  
In addition,   one has to put   $\delta$ in the propagators    
to zero ($\delta=0$)  for  $NN$  intermediate  states  
while  for a $N \Delta$ intermediate  
state  $\delta= m_{\Delta}-m_N$.

In case of a deuteron in the final state, the corresponding $\mathcal M$ matrices
should be replaced by the deuteron wave functions according to
\begin{eqnarray}
\maM^\mathrm{FSI}[jls,1]&=&  
\frac{1}{\sqrt{2M_N}}
\sum_{l^{\prime\prime}}\int\frac{d^3q}{(2\pi)^3}\ \tilde\maM^\mathrm{prod}[jls,1l''s''](p,q)
i^{l''}\psi^{l''}(q),
\label{dfsi}
\end{eqnarray}
where $\psi^{l^{\prime\prime}}(q)$ are the deuteron wave functions corresponding 
to the angular momentum $l^{\prime\prime}$, normalized by the condition 
\begin{equation}
	\int \frac{d^3 q}{(2\pi)^3}\left( (\psi^0(q))^2+ (\psi^2(q))^2\right)=1.
\end{equation}
Thus, the two-nucleon propagator for 
the deuteron in the final state is absorbed in the wave functions and the normalization 
has changed. 
Analogous expressions can be written down for the deuteron in the initial state.
Note that in the case of the deuteron in the initial  or final state
the N$^2$LO production operator derived in this  paper appears 
only as the building block for the calculation of the
$\mathcal{M}^\text{FSI}$, $\mathcal{M}^\text{ISI}$ and 
$\mathcal{M}^\text{ISI+FSI}$ amplitudes 
according to Eqs.~(\ref{fsi})--(\ref{fsiisi}) and (\ref{dfsi}),
respectively. They do not contribute independently because then there are no free 
nucleons in the initial or final state.

%%%%%%%%%%%%%%%%%%%%%%%%%%%%%%%%%%%%%%%%%%%%%%%%%%%%%%%%%%%%%%%%%%%%%%%%%%%
\section{The pion-nucleon Lagrangian}
\label{lagrang}
%%%%%%%%%%%%%%%%%%%%%%%%%%%%%%%%%%%%%%%%%%%%%%%%%%%%%%%%%%%%%%%%%%%%%%%%%%%%%

The leading order pion nucleon Lagrangian has form
\begin{equation} {\cal L}_{\pi N}^{(1)}  = \bar{N}
 \left( i v \cdot D  + {g}_A S \cdot u \right) N.
 \label{lagr} 
\end{equation}
The  pion fields is contained non-linearly in the $u$-field which
in the sigma gauge is:
\begin{equation}
	u(\boldpi) 
	= \sqrt{U(\boldpi)} 
	= 1 + i\frac{\boldtau \cdot \boldpi}{2 f_\pi} - \frac{\boldpi^2}{8 f_\pi^2} 
	    + i \frac{(\boldtau \cdot \boldpi)^3}{16 f_\pi^3} + \cdots.
\end{equation}
Furthermore,  the chiral Lagrangian contains the derivative of the $u$-field 
and the chiral covariant derivative $D_\mu$ defined as:
\begin{eqnarray}
u_\mu & = & i\,\left(\partial_\mu u\,u^\dagger + u^\dagger\partial_\mu u \right), 
\\
D_\mu  & = & \partial_\mu\, + \Gamma_\mu ,
\\
\Gamma_\mu & = & \frac{1}{2}\,[u^\dagger,\partial_\mu u].
\end{eqnarray}
The heavy baryon formalism involves the covariant spin-operator $S_\mu$ 
and the four-velocity $v^\mu$ where
\begin{equation} 
S_\mu = \frac{i}{2} \gamma_5 \sigma_{\mu \nu} v^\nu \, , \quad 
S \cdot v = 0 \, , \quad 
\lbrace S_\mu , S_\nu \rbrace 
     = \frac{1}{2} \left( v_\mu v_\nu - g_{\mu \nu} \right) \, . 
\label{spin}
 \end{equation}
 In  four-dimensional space-time, the commutator of two spin-operators
 can be simplified to
 $[S_\mu , S_\nu] = i \epsilon_{\mu \nu \gamma \delta} v^\gamma S^\delta$,
where we use the convention $\epsilon^{0123} = -1$.

The next order Lagrangian has two derivatives or one $m_\pi^2$ insertion:
\begin{eqnarray} {\cal L}_{\pi N}^{(2)} & = & \bar{N}
 \left\{ \frac{1}{2m_N} (v \cdot D)^2 - \frac{1}{2m_N} D\cdot D  - 
\frac{ i\, g_A}{2m_N} \left\{ S \cdot D, \, v\cdot u \right\} 
+c_1 \langle \chi_+ \rangle 
\right. \nonumber \\ && 
\left.
+\left(c_2-\frac{g_A^2}{8m_N}\right) (v\cdot u)^2 +c_3 u\cdot u 
+\left(c_4+\frac{1}{4m_N}\right)\left[S^\mu , S^\nu \right] u_\mu u_\nu 
\right\} N 
+ \cdots  \, , 
\label{lagr2} 
\end{eqnarray}
where $\chi_+ =u^\dagger \chi u^\dagger + u \chi^\dagger u$ 
($\chi = m_\pi^2$  up  to  isospin violating  corrections) and 
where we have only written those terms relevant for this paper, 
in addition `$\langle \ldots \rangle$' denotes the trace in flavor space.

The relevant third order Lagrangian takes the form
\begin{eqnarray} \label{lpin3f}
{\cal L}_{\pi N}^{(3)}
& = & \bar{N} 
{\cal O}^{(3)}_{\rm fixed} 
N \,\,\, ,
\end{eqnarray}
where 
\begin{eqnarray}
{\cal O}^{(3)}_{\rm fixed} 
& = &
\frac{g_A}{8m_N^2}\,[D^\mu,[D_\mu,S\!\cdot\! u]]
-i\,\frac{1}{4m_N^2}\left(v\!\cdot\! D\right)^3
\nonumber \\%[0.5em]
&& - 
\frac{g_A}{4m_N^2}\,v\cdot\!\!\stackrel{\leftarrow}{D} S\!\cdot\!u\,v\!\cdot\! D
+\frac{1}{8m_N^2}\left(i\,D^2\, v\!\cdot\! D + \mbox{h.c.}\right)
\nonumber \\%[0.5em]
&& - 
\frac{g_A}{4m_N^2}\left(\{S\!\cdot\! D,v\!\cdot\! u\}\,v\!\cdot\! D + \mbox{h.c.}\right)
+\frac{3g_A^2}{64m_N^2}\left(i\langle(v\!\cdot\!u)^2\rangle\,v\!\cdot\!D + \mbox{h.c.}\right)
\nonumber \\%[0.5em]
&& + 
\frac{1}{32m_N^2}\left(\epsilon^{\mu\nu\alpha\beta}v_\alpha S_\beta
[u_\mu,u_\nu]\,v\!\cdot\! D + \mbox{h.c.}\right)
\nonumber \\%[0.5em]
%\\[0.5em]
&&-  
\frac{g_A}{8m_N^2}\left(S\!\cdot\! u\,D^2 + \mbox{h.c.}\right)
-\frac{g_A}{4m_N^2}\left(S\cdot\!\!\stackrel{\leftarrow}{D} u\!\cdot\! D + \mbox{h.c.}\right)
\nonumber \\%[0.5em]
&& + 
\frac{1+g_A^2+8m_N\,c_4}{16m_N^2}\left(\epsilon^{\mu\nu\alpha\beta}v_\alpha S_\beta
[u_\mu,v\!\cdot\! u] D_\nu + \mbox{h.c.}\right)
\nonumber \\%[0.5em]
&&
-\frac{g_A^2}{16m_N^2}\left(i\,v\!\cdot\!u\,u\!\cdot\!D  + \mbox{h.c.}\right)
\nonumber \\%[0.5em]
&& + 
i\,\frac{1+8m_N\,c_4}{32m_N^2}\,[v\!\cdot\! u,[D^\mu,u_\mu]]
+\frac{c_2}{2m_N}\left(i\langle v\!\cdot\!u\,u_\mu\rangle D^\mu + \mbox{h.c.}\right) + \cdots. 
\end{eqnarray}

%%%%%%%%%%%%%%%%%%%%%%%%%%%%%%%%%%%%%%%%%%%%%%%%%%%%%%%%%%%%%%%%%%%%%%%%%%%%%%%%%%%%%%%%%%%%%

%%%%%%%%%%%%%%%%%%%%%%%%%%%%%%%%%%%%%%%%%%%%%%%%%%%%%%%%%%%%%%%%%%%%%%%%%%%%%%%%
\section{Basic integrals}
\label{sec:appbasicint}
%%%%%%%%%%%%%%%%%%%%%%%%%%%%%%%%%%%%%%%%%%%%%%%%%%%%%%%%%%%%%%%%%%%%%%%%%%%%%%%%%%

\subsection{Definitions and analytic expressions for  various integrals} 
%%%%%%%%%%%%%%%%%%%%%%%%%%%%%%%%%%%%%%%%%%%%%%%%%%%%%%%%%%

In this subsection we give the explicit definitions of the 
common dimensionless loop integrals used in this work. The first integral 
$J_{\pi \Delta}= \mu^\epsilon J_0(-\delta )$ where $\mu$  is the  dimension-regularization scale and the  integral $J_0(-\delta )$ is defined 
in Ref.~\cite{ulfbible}. 
\begin{eqnarray}
	\frac{1}{\delta}\jpid (\delta)
		   &=& \frac{\mu^\epsilon}{i \delta} \int \frac{d^{4- \epsilon}l}{(2\pi)^{4-\epsilon}}
	         \frac{1}{(l^2-\mpi^2+i0)( - v \cdot l - \delta +i0)}	
	         \nonumber 
\\
	       &=& 4 L + \frac{(-2)}{(4\pi)^2} \left[ -1 + \log \left( \frac{\mu^2}{\mpi^2} \right) \right] 
	       \nonumber 
\\
	       &&+ \frac{4}{(4\pi)^2} 
	           \left[ -1 + 
	                \frac{\sqrt{1-y-i0}}{\sqrt{y}} 
	                \left[ 
	                      -\frac{\pi}{2} + \arctan \left( \frac{\sqrt{y}}{\sqrt{1-y-i0}} \right)
	                \right] 
	           \right],
		\label{JpiD}
\\
	\ipipi (k_1^2)
		&=& \frac{\mu^\epsilon}{i} \int \frac{d^{4- \epsilon}l}{(2\pi)^{4-\epsilon}}
		\frac{1}{(l^2-\mpi^2+i0)((l+k_1)^2 - \mpi^2 +i0)} 
		\nonumber
\\
		&=& - 2 L - \frac{1}{(4 \pi)^2} \left [ \log \left( \frac{\mpi^2}{\mu^2} \right)
		- 1 + 2 F_1 \left(\frac{k_1^2}{\mpi^2}\right) \right],
		\label{Ipipi}
\end{eqnarray}
where  
\begin{eqnarray}
\newcommand{\myvar}{x}
F_1 (\myvar) = \frac{\sqrt{4-\myvar-i0}}{\sqrt{\myvar}}
\arctan \left( \frac{\sqrt{\myvar}}{\sqrt{4-\myvar-i0}} \right)\,,
\label{eq:F1} 
\end{eqnarray}
\begin{equation}
	L = \frac{1}{(4\pi)^2} \left[ - \frac{1}{\epsilon} + \frac12 \left( \gamma_E -1 -\log (4\pi) \right)  \right],
\end{equation}
and the variables $x$, $y$ are defined via $x= k_1^2 / \mpi^2$, $y = \delta^2 / \mpi^2 $.

Further, the integrals in Eqs.~(\ref{JpipiD}) and (\ref{JpipiND}) 
can be reduced to simple 
one-dimensional integrals which can be calculated numerically.
\begin{eqnarray}
	\intT &=& \delta \frac{\mu^\epsilon}{i} \int \frac{d^{4- \epsilon}l}{(2\pi)^{4-\epsilon}}
         \frac{1}{(l^2-\mpi^2+i0)((l+k_1)^2 - \mpi^2 +i0)( - v \cdot l - \delta +i0)},
         \label{JpipiD}
\\
	\intL &=& k_1^2 \frac{\mu^\epsilon}{i} \int \frac{d^{4- \epsilon}l}{(2\pi)^{4-\epsilon}}
         \Big[ \frac{1}{(l^2-\mpi^2+i0)((l+k_1)^2 - \mpi^2 +i0)} 
         \nonumber
\\ 
         && \times \frac{1}{( - v \cdot l  +i0) ( - v \cdot l - \delta +i0)} \Big].
         \label{JpipiND}
\end{eqnarray}

It is also  convenient  to define  finite, scale-independent parts of
$\jpid$ and $\ipipi$ in which  the  divergency  $L $ and the $\log
(\mpi/\mu)$   terms  are removed.  The  finite  contributions 
$\jpidfsi$  and $ \ipipifsi$  will be used in the subsequent section.

%%%%%%%%%%%%%%%%%%%%%%%%%%%%%%%%%%%%%%%%%%%%%%%%%%%%%%%%%%%%%%%
\begin{eqnarray}
    \ipipi
        &=&  -2 L - \frac{1}{(4\pi)^2} \log \left( \frac{\mpi^2}{\mu^2} \right) + \ipipifsi,
\\
    \ipipifsi 
        &=&  
		\frac{1}{(4\pi)^2}
		\left( 
		1-  2 \frac{\sqrt{4-x-i0}}{\sqrt{x}} \arctan \left( \frac{\sqrt{x}}{\sqrt{4-x-i0}} \right)
		\right),
		\label{ipipifsi}
\\    
    \frac{1}{\delta}\jpid 
           &=& 4 L +
            \frac{2}{(4\pi)^2} 
           \log \left( \frac{\mpi^2}{\mu^2} \right)   + \frac{1}{\delta} \jpidfsi,
\\
    \frac{1}{\delta} \jpidfsi
           &=&  \frac{4}{(4\pi)^2} 
               \left[ -\frac{1}{2} + 
                    \frac{\sqrt{1-y-i0}}{\sqrt{y}} 
                    \left[ 
                          -\frac{\pi}{2} + \arctan \left( \frac{\sqrt{y}}{\sqrt{1-y-i0}} \right)
                    \right] 
               \right]. 
\end{eqnarray}

From the expressions above it is easy  to obtain  the  important
relation,  which  is  used  in the analysis of the  integral
combinations,  relevant for our study
\begin{eqnarray}
    \ipipi+\frac{1}{2\delta}\jpid = \ipipifsi + \frac{1}{2\delta}
    \jpidfsi.
\label{intfinite}
\end{eqnarray}

%%%%%%%%%%%%%%%%%%%%%%%%%%%%%%%%%%%%%%%%%%%%%%%%%%%%%%%%%%%%%%%%%%%%%%%%%%%%%%%%
%%%%%%%%%%%%%%%%%%%%%%%%%%%%%%%%%%%%%%%%%%%%%%%%%%%%%%%%%%%%%%%%%%%%%%%%%%%%%%%%%%%
\subsection{Combinations of basic integrals  in  the  limit  $\delta \to \infty$}
\label{intdecoupl}
%%%%%%%%%%%%%%%%%%%%%%%%%%%%%%%%%%%%%%%%%%%%%%%%%%%%%%%%%%%%%%%%%%%%%%%%%%%%%%%

In  this  subsection  we  discuss  the behavior of   the  integral
combinations, see Eqs.~(\ref{intcomb1})--(\ref{intcomb3}), 
relevant for   loop-diagrams considered in this work.  
While the integrals $\ipipi$ and $\jpid$  have analytic  expressions
for  any  $\delta$,  the integral $\jpipid$  can be done analytically
only  in the limit  $\delta \to \infty$.
Using  the  dispersive analysis,  
one  finds the asymptotic expression for  $\delta \jpipid$  when $\delta \to \infty$  
\begin{eqnarray}
	(4 \pi)^2 \delta \jpipid &=& 
	2 \log \left[\frac{\mpi}{2\delta }\right]
	-1
	-(4 \pi)^2 \ipipifsi
	+\frac{1}{36 \delta ^2}
	\Big(
	(36 \mpi^2-6 k_1^2) \log \left[\frac{\mpi}{2\delta }\right]
	\nonumber \\
	&&-k_1^2+6 \mpi^2+(3 k_1^2 -12 \mpi^2 ) (4 \pi)^2 \ipipifsi
		\Big)
	+O\left(\frac{1}{\delta^3 }\right).
\end{eqnarray}
Note that to get  the MCS terms  relevant at  \NNLO{},   one ignores       
$\mpi^2$ compared  to $k_1^2 $ in the last  two lines of   the
expression  above.

Using the expression above,  Eq.~(\ref{intfinite})  and the  expansion  of  $\jpidfsi$ for 
large $\delta$
\begin{eqnarray}
    \frac{(4\pi)^2 }{\delta} \jpidfsi = 
    - 4 \log\left[\frac{\mpi}{2\delta }\right] -2
    +\frac{\mpi^2}{\delta ^2}
    \left( 
        2 \log \left[\frac{\mpi}{2\delta }\right] -1
     \right) 
    +O\left(\frac{1}{\delta^4 }\right),
\end{eqnarray}
one  obtains
\begin{eqnarray}
		\left( \intI + \intT + \frac12 \intJ + \intC{2} \right) 
		= O\left(\frac{1}{\delta^2 }\right).
\end{eqnarray}
Thus,  this  combination  vanishes in the  limit  $\delta \to \infty$.

Analogously,  one finds 

\begin{eqnarray}
		&&\frac{\delta^2}{k_1^2} \left( \intI + \intT + \frac12 \intJ + \intC{2}\right) 
	    	- \frac{1}{12} \left(  \intI + \frac12 \intJ + \frac13 \intC{2} \right) 
	    	\nonumber \\
	    	&&= \frac{\mpi^2}{3 k_1^2} \left( 6  \log\left[\frac{\mpi}{2\delta }\right]-1- \ipipifsi \right) 
	    	+O\left(\frac{1}{\delta^2 }\right).
\end{eqnarray}
This combination vanishes in the  limit  $\delta \to \infty$ up to
higher order terms. To make it vanishing also at higher  order one  would
need to extend the                 
calculation and keep  the so far  neglected
higher order terms.

Finally, the integral $\jpipind$  obviously vanishes at large $\delta$ 
\begin{eqnarray}
			\jpipind= \frac{1}{\delta} (\jpipid-\jpipin),
\end{eqnarray}
where  $\jpipin=\jpipid(\delta=0)$.

 %%%%%%%%%%%%%%%%%%%%%%%%%%%%%%%%%%%%%%%%%%%%%%%%%%%%%%%%%%%%%%%%%%%%%%%%%%%%%%%%%


\begin{thebibliography}{99}
%%%%%%%%%%%%%%%%%%%%%%%%%%%%%%%%%%%%%%%%%%%%%%%%%%%%%%%%%%%%%%%%%%%%%%%%%%%%%%%%%%


\bibitem{cohen}
  T.~D.~Cohen, J.~L.~Friar, G.~A.~Miller and U.~van Kolck,
  %``The p p ---> p p pi0 reaction near threshold: A Chiral power counting approach,''
  Phys.\ Rev.\ C\ {\bf 53},  2661 (1996).
  %[nucl-th/9512036].
  %%CITATION = PHRVA,C53,2661;%%
  

 \bibitem{park}
  B.~Y.~Park, F.~Myhrer, J.~R.~Morones, T.~Meissner and K.~Kubodera,
  %``Chiral perturbation approach to the p p ---> p p pi0 reaction near threshold,''
  Phys.\ Rev.\ C\ {\bf 53},  1519 (1996).
  %[nucl-th/9512023].
  %%CITATION = PHRVA,C53,1519;%%
  
\bibitem{hanhart04}
  C.~Hanhart,
  %``Meson production in nucleon-nucleon collisions close to the threshold,''
  Phys.\ Rept.\ \ {\bf 397},  155 (2004).
  %[hep-ph/0311341].
  %%CITATION = PRPLC,397,155;%%
  
  
\bibitem{BHM}
 V.~Baru,  C.~Hanhart,  and  F.~Myhrer, 
 ``Effective field theory  calculations for  $NN\to NN\pi$'',   in  preparation 

  
\bibitem{Bernardrev}
V.~Bernard,
  %``Chiral Perturbation Theory and Baryon Properties,''
  Prog.\ Part.\ Nucl.\ Phys.\  {\bf 60}, 82 (2008).
 % [arXiv:0706.0312 [hep-ph]].
  %%CITATION = ARXIV:0706.0312;%%

\bibitem{NN3}
  E.~Epelbaum, H.-W.~Hammer and U.-G.~Mei\ss ner,
  %``Modern Theory of Nuclear Forces,''
  Rev.\ Mod.\ Phys.\  {\bf 81}, 1773 (2009).
 % [arXiv:0811.1338 [nucl-th]].
  %%CITATION = ARXIV:0811.1338;%%}  // %, Hammer H. W.  and Mei{\ss}ner U.- G.  } 



\bibitem{sato}
  T.~Sato, T.~S.~H.~Lee, F.~Myhrer and K.~Kubodera,
  %``Chiral perturbation theory and the p p ---> p p pi0 reaction near threshold,''
  Phys.\ Rev.\ C\ {\bf 56},  1246 (1997).
  %[nucl-th/9704003].
  %%CITATION = PHRVA,C56,1246;%%

\bibitem{unserd}
  C.~Hanhart, J.~Haidenbauer, M.~Hoffmann, U.-G.~Mei{\ss}ner and J.~Speth,
  %``The Reactions p p ---> p p pi0 and p p ---> d pi+ at threshold: The Role of the isoscalar pi N scattering amplitude,''
  Phys.\ Lett.\ B\ {\bf 424},  8 (1998).
  %[nucl-th/9707029].
  %%CITATION = PHLTA,B424,8;%%

\bibitem{DKMS}
  V.~Dmitrasinovic, K.~Kubodera, F.~Myhrer and T.~Sato,
  %``A Next-to-next-to leading order p p ---> p p pi0 transition operator in chiral perturbation theory,''
  Phys.\ Lett.\ B\ {\bf 465},  43 (1999).
%  [nucl-th/9902048].
  %%CITATION = PHLTA,B465,43;%% 

\bibitem{Ando}
  S.~Ando, T.-S.~Park and D.-P.~Min,
  %``Threshold p p ---> p p pi0 up to one loop accuracy,''
  Phys.\ Lett.\ B\ {\bf 509},  253 (2001).
%  [nucl-th/0003004].
  %%CITATION = PHLTA,B509,253;%%
  
  

\bibitem{novel}
  V.~Bernard, N.~Kaiser and U.-G.~Mei{\ss}ner,
  %``Novel approach to pion and eta production in proton proton collisions,''
  Eur.\ Phys.\ J.\ A\ {\bf 4},  259 (1999).
  %[nucl-th/9806013].
  %%CITATION = EPHJA,A4,259;%%


\bibitem{rocha}
  C.~A.~da Rocha, G.~A.~Miller and U.~van Kolck,
  %``The N N ---> N N pi+ reaction near threshold in a chiral power counting approach,''
  Phys.\ Rev.\ C\ {\bf 61},  034613 (2000).
  %[nucl-th/9904031].
  %%CITATION = PHRVA,C61,034613;%%
  
\bibitem{HanKai}
  C.~Hanhart and N.~Kaiser,
  %``Complete next-to-leading order calculation for pion production in nucleon-nucleon collisions at threshold,''
  Phys.\ Rev.\ C\ {\bf 66},  054005 (2002) .
  %[nucl-th/0208050].
  %%CITATION = PHRVA,C66,054005;%%


\bibitem{towards}
  V.~Lensky, V.~Baru, J.~Haidenbauer, C.~Hanhart, A.~E.~Kudryavtsev and U.-G.~Mei{\ss}ner,
  %``Towards a field theoretic understanding of NN ---> NN pi,''
  Eur.\ Phys.\ J.\ A\ {\bf 27},  37 (2006).
  %[nucl-th/0511054].
  %%CITATION = EPHJA,A27,37;%%

\bibitem{ksmk09}
  Y.~Kim, T.~Sato, F.~Myhrer and K.~Kubodera,
  %``Two-pion-exchange and other higher-order contributions to the pp ---> pp pi0 reaction,''
  Phys.\ Rev.\ C\ {\bf 80},  015206 (2009).
  %[arXiv:0810.2774 [nucl-th]].
  %%CITATION = PHRVA,C80,015206;%%

\bibitem{NNLOswave} 
  A.~A.~Filin, V.~Baru, E.~Epelbaum, H.~Krebs, C.~Hanhart, A.~E.~Kudryavtsev and F.~Myhrer,
  %``Pion production in nucleon-nucleon collisions in chiral effective field theory: next-to-next-to-leading order contributions,''
  Phys.\ Rev.\ C {\bf 85}, 054001 (2012).
%  [arXiv:1201.4331 [nucl-th]].
  %%CITATION = ARXIV:1201.4331;%%



\bibitem{subloops}
  C.~Hanhart and A.~Wirzba,
  %``Remarks on N N ---> N N pi beyond leading order,''
  Phys.\ Lett.\ B\ {\bf 650},  354 (2007).
  %[nucl-th/0703012 [NUCL-TH]].
  %%CITATION = PHLTA,B650,354;%%

\bibitem{BM}
  D.~R.~Bolton and G.~A.~Miller,
  %``Impulse approximation in nuclear pion production reactions: absence of a one-body operator,''
  Phys.\ Rev.\ C\ {\bf 83},  064003 (2011).
  %[arXiv:1008.3378 [nucl-th]].
  %%CITATION = PHRVA,C83,064003;%%


\bibitem{ch3body}
  C.~Hanhart, U.~van Kolck and G.~A.~Miller,
  %``Chiral three nucleon forces from p wave pion production,''
  Phys.\ Rev.\ Lett.\ \ {\bf 85},  2905 (2000).
  %[nucl-th/0004033].
  %%CITATION = PRLTA,85,2905;%%


\bibitem{newpwave}
  V.~Baru, E.~Epelbaum, J.~Haidenbauer, C.~Hanhart, A.~E.~Kudryavtsev, V.~Lensky and U.-G.~Mei{\ss}ner,
  %``p-wave pion production from nucleon-nucleon collisions,''
  Phys.\ Rev.\ C\ {\bf 80},  044003 (2009).
  %[arXiv:0907.3911 [nucl-th]].
  %%CITATION = PHRVA,C80,044003;%% 

\bibitem{nakamura}
S.~X.~Nakamura,
  %``Bridging over p-wave pi-production and weak processes in few-nucleon systems with chiral perturbation theory,''
  Phys.\ Rev.\ C {\bf 77}, 054001 (2008).
%  [arXiv:0709.1239 [nucl-th]].
  %%CITATION = ARXIV:0709.1239;%%





\bibitem{KNM}  U.~van Kolck, J.~A.~Niskanen and G.~A.~Miller,
  %``Charge symmetry violation in p n ---> d pi0 as a test of chiral effective field theory,''
  Phys.\ Lett.\ B {\bf 493}, 65 (2000).
%  [nucl-th/0006042].
  %%CITATION = NUCL-TH/0006042;%%


\bibitem{weCSB}  
 A.A.~Filin,  V.~Baru, E.~Epelbaum, J.~Haidenbauer, C.~Hanhart, 
A.~E.~Kudryavtsev and U.-G.~Mei{\ss}ner,
  %``Extraction of the strong neutron-proton mass difference from the 
  % charge symmetry breaking in pn ---> d pi0,''
  Phys.\ Lett.\ B {\bf 681}, 423 (2009).
%  [arXiv:0907.4671 [nucl-th]].
  %%CITATION = ARXIV:0907.4671;%%

\bibitem{Bolton} 
D.~R.~Bolton and G.~A.~Miller,
  %``Charge Symmetry Breaking in the n p ---> d pi0 reaction,''
  Phys.\ Rev.\ C {\bf 81}, 014001 (2010).
%  [arXiv:0907.0254 [nucl-th]].
  %%CITATION = ARXIV:0907.0254;%%


\bibitem{koltun}
  D.~S.~Koltun and A.~Reitan,
  %``Production and absorption of S-wave pions at low energy by two nucleons,''
  Phys.\ Rev.\ \ {\bf 141}, 1413 (1966).
  %%CITATION = PHRVA,141,1413;%%
  
\bibitem{Lee}
  T.~S.~H.~Lee and D.~O.~Riska,
  %``Short range exchange contributions to the cross-section for p p ---> p p pi0 near threshold,''
  Phys.\ Rev.\ Lett.\ \ {\bf 70}, 2237 (1993).
  %%CITATION = PRLTA,70,2237;%%


\bibitem{HGM}
  C.~J.~Horowitz, H.~O.~Meyer and D.~K.~Griegel,
  %``Role of heavy meson exchange in pion production near threshold,''
  Phys.\ Rev.\ C\ {\bf 49},  1337 (1994).
  %[nucl-th/9304004].
  %%CITATION = PHRVA,C49,1337;%%

\bibitem{Hpipl}
  C.~J.~Horowitz,
  %``Role of heavy meson exchange in near threshold N N ---> d pi,''
  Phys.\ Rev.\ C\ {\bf 48},  2920  (1993).
  %%CITATION = PHRVA,C48,2920;%%

\bibitem{jounicomment}
%\bibitem{nucl-th/9502015}
  J.~A.~Niskanen,
  %``Comment on 'Role of heavy meson exchange in near threshold N N ---> d pi',''
  Phys.\ Rev.\ C\ {\bf 53},  526, (1996).
  %[nucl-th/9502015].
  %%CITATION = PHRVA,C53,526;%%

\bibitem{eulogio}
%\bibitem{nucl-th/9503019}
  E.~Hernandez and E.~Oset,
  %``Off-shell pi N amplitude and the p p ---> p p pi0 reaction near threshold,''
  Phys.\ Lett.\ B\ {\bf 350}, 158 (1995).
  %[nucl-th/9503019].
  %%CITATION = PHLTA,B350,158;%%

\bibitem{unsers}
%\bibitem{nucl-th/9508005}
  C.~Hanhart, J.~Haidenbauer, A.~Reuber, C.~Schutz and J.~Speth,
  %``The Reaction p p ---> p p pi0 near threshold,''
  Phys.\ Lett.\ B\ {\bf 358},  21 (1995).
  %[nucl-th/9508005].
  %%CITATION = PHLTA,B358,21;%%

\bibitem{unserdelta}
 C.~Hanhart, J.~Haidenbauer, O.~Krehl and J.~Speth,
  %``Role of the Delta isobar in the reaction N N ---> N N pi near threshold,''
  Phys.\ Lett.\ B {\bf 444},  25 (1998).


\bibitem{compton} 
  V.~Pascalutsa and D.~R.~Phillips,
  %``Effective theory of the delta(1232) in Compton scattering off the nucleon,''
  Phys.\ Rev.\ C {\bf 67},  055202  (2003).
%  [nucl-th/0212024].
  %%CITATION = NUCL-TH/0212024;%%


\bibitem{NNpiMenu}
  V.~Baru, J.~Haidenbauer, C.~Hanhart, A.~E.~Kudryavtsev, V.~Lensky and U.-G.~Mei{\ss}ner,
  %``Progress in NN ---> NN pi,''
 in {\it Proceedings of 11-th International
Conference on Meson-Nucleon Physics and the
Structure of the Nucleon (MENU 2007), J\" ulich, Germany},   eConfC\ {\bf 070910} 128 (2007);
  [arXiv:0711.2748 [nucl-th]].
  %%CITATION = ECONF,C070910,128;%%



\bibitem{OvK}
  C.~Ord\'o\~nez, L.~Ray and U.~van Kolck,
  %``The Two nucleon potential from chiral Lagrangians,''
  Phys.\ Rev.\ C\ {\bf 53},  2086  (1996).
  %[hep-ph/9511380].
  %%CITATION = PHRVA,C53,2086;%%


\bibitem{ulfbible}
  V.~Bernard, N.~Kaiser and U.-G.~Mei{\ss}ner,
  %``Chiral dynamics in nucleons and nuclei,''
  Int.\ J.\ Mod.\ Phys.\ E\ {\bf 4}, 193  (1995).
  %[hep-ph/9501384].
  %%CITATION = IMPAE,E4,193;%%


\bibitem{Fettes}
N. Fettes, U.-G. Mei{\ss}ner and S. Steininger, 
Nucl. Phys. A {\bf 640}, 199 (1998); \\
N. Fettes, Ph.D. thesis, Bonn University, 2000.

\bibitem{Fettes2}
  N.~Fettes, U.-G.~Mei{\ss}ner, M.~Mojzis and S.~Steininger,
  %``The Chiral effective pion nucleon Lagrangian of order p**4,''
  Annals Phys.\  {\bf 283},  273   (2000) [Erratum-ibid.\  {\bf 288},  249  (2001)].
  %%CITATION = HEP-PH/0001308;%%




\bibitem{Gardestig}
  A.~G{\aa}rdestig, D.~R.~Phillips and C.~Elster,
  %``The Near threshold N N ---> d pi reaction in chiral perturbation theory,''
  Phys.\ Rev.\ C\ {\bf 73},  024002  (2006).
  %[nucl-th/0511042].
  %%CITATION = PHRVA,C73,024002;%%



  \bibitem{Hemmert}
T.~R.~Hemmert,   PhD-thesis,  University of  Massachusetts  Amherst  (1999).



\bibitem{Hemmert:1997ye}
  T.~R.~Hemmert, B.~R.~Holstein and J.~Kambor,
  %``Chiral Lagrangians and delta(1232) interactions: Formalism,''
  J.\ Phys.\ G {\bf 24}, 1831  (1998).
%  [hep-ph/9712496].
  %%CITATION = HEP-PH/9712496;%%
  %214 citations counted in INSPIRE as of 12 Jul 2013

\bibitem{FettesDel} 
  N.~Fettes and U.~G.~Mei{\ss}ner,
  %``Pion - nucleon scattering in an effective chiral field theory with explicit spin 3/2 fields,''
  Nucl.\ Phys.\ A {\bf 679}, 629 (2001).
%  [hep-ph/0006299].
  %%CITATION = HEP-PH/0006299;%%

\bibitem{Kaiser1998}
N. Kaiser, S. Gerstendorfer, W. Weise, Nucl. Phys. A {\bf 637}, 395 (1998).


\bibitem{wein90}
  S.~Weinberg,
  %``Nuclear forces from chiral Lagrangians,''
  Phys.\ Lett.\ B {\bf 251},  288  (1990).
  %%CITATION = PHLTA,B251,288;%%
  
\bibitem{weinberg1991} 
  S.~Weinberg,
  %``Effective chiral Lagrangians for nucleon - pion interactions and nuclear forces,''
  Nucl.\ Phys.\ B {\bf 363},  3 (1991).
  %%CITATION = NUPHA,B363,3;%% 



\bibitem{CCF}
  J.~Haidenbauer, K.~Holinde and M.~B.~Johnson,
  %``A Coupled channel potential for nucleons and deltas,''
  Phys.\ Rev.\ C {\bf 48},  2190  (1993).
  %%CITATION = PHRVA,C48,2190;%%
  
\bibitem{CDBonn}
  R.~Machleidt,
  %``The High precision, charge dependent Bonn nucleon-nucleon potential (CD-Bonn),''
  Phys.\ Rev.\ C {\bf 63}, 024001  (2001).
%  [nucl-th/0006014].

\bibitem{Bernard:1998gv}
  V.~Bernard, H.~W.~Fearing, T.~R.~Hemmert and U.-G.~Mei{\ss}ner,
  %``The form-factors of the nucleon at small momentum transfer,''
  Nucl.\ Phys.\ A {\bf 635},  121 (1998) 
   [Erratum-ibid.\ A {\bf 642},  563  (1998)]
   [Nucl.\ Phys.\ A {\bf 642}, 563  (1998)].
%  [hep-ph/9801297].
  %%CITATION = HEP-PH/9801297;%%
  %129 citations counted in INSPIRE as of 12 Jul 2013

\bibitem{Appelquist} 
T. Appelquist and J. Carazzone, Phys. Rev. D {\bf 11},  2856 (1975);
Lowell S. Brown, Phys. Rev. D {\bf 39}, 3084 (1989). 




\bibitem{charmloops}
 M.~Cleven, F.-K.~Guo, C.~Hanhart and U.-G.~Mei{\ss}ner,
  %``Bound state nature of the exotic $Z_b$ states,''
  Eur.\ Phys.\ J.\ A {\bf 47},  120  (2011).
%  [arXiv:1107.0254 [hep-ph]].



\end{thebibliography}
\end{document}